%% file: example_paper.tex
\documentclass{article}

\usepackage{microtype}
\usepackage{graphicx}
\usepackage{subcaption}
\usepackage{booktabs} % for professional tables
\usepackage{amsthm}
\usepackage{amsmath}
\usepackage{amssymb}
\usepackage{mathtools}
\usepackage{siunitx} 
\usepackage{multirow}
\usepackage{colortbl}
\usepackage{makecell}

\usepackage{chngcntr}  % table and fig number as section.

\usepackage{hyperref}

\usepackage[preprint]{icml2026}

\definecolor{bestcolor}{RGB}{255, 224, 224}
\definecolor{secondcolor}{RGB}{224, 224, 255}
\definecolor{normalcolor}{RGB}{255, 255, 255} % White for normal cells

\newcommand{\perfalign}[3]{% #1: value, #2: symbol, #3: std
    \begin{tabular}{@{}r@{\,}l@{}}
        #1 & #2\ #3
    \end{tabular}
}
\newcommand{\perfalignplain}[2]{% #1: value, #2: std (for DCFormer)
    \begin{tabular}{@{}r@{\ }l@{}}
        #1 & #2
    \end{tabular}
}

\newcommand{\best}[1]{\cellcolor{bestcolor}\perfalign{#1}}
\newcommand{\secondd}[1]{\cellcolor{secondcolor}\perfalign{#1}}
\newcommand{\normal}[1]{\cellcolor{normalcolor}\perfalign{#1}}
\newcommand{\bestplain}[1]{\cellcolor{bestcolor}\perfalignplain{#1}}
\newcommand{\secondplain}[1]{\cellcolor{secondcolor}\perfalignplain{#1}}

\usepackage[capitalize,noabbrev]{cleveref}

\theoremstyle{plain}

\theoremstyle{definition}

\theoremstyle{remark}

\usepackage[textsize=tiny]{todonotes}

\icmltitlerunning{Hyperspectral Image Fusion with Spectral-Band and Fusion-Scale Agnosticism}

\begin{document}

\twocolumn[
  \icmltitle{Hyperspectral Image Fusion with Spectral-Band and Fusion-Scale Agnosticism}



  \icmlsetsymbol{equal}{*}

  \begin{icmlauthorlist}
    \icmlauthor{Yu-Jie Liang}{equal,uestcCS}
    \icmlauthor{Zihan Cao}{equal,uestcMath}
    \icmlauthor{Liang-Jian Deng}{uestcMath,multiHazardEarlyWarningDeng}
    \icmlauthor{Yang Yang}{uestcCS}
    \icmlauthor{Malu Zhang}{uestcCS}
  \end{icmlauthorlist}

  \icmlaffiliation{uestcMath}{University of Electronic Science and Technology of China, School of Mathematical Sciences, Chengdu, China}
  
  \icmlaffiliation{multiHazardEarlyWarningDeng}{University of Electronic Science and Technology of China, Multi-Hazard Early Warning Key Laboratory of Sichuan Province, Chengdu, China}

  \icmlaffiliation{uestcCS}{University of Electronic Science and Technology of China, School of Computer Science and Engineering (School of Cyber Security), Chengdu, China}

  \icmlcorrespondingauthor{Malu Zhang}{maluzhang@uestc.edu.cn}
  
  \icmlkeywords{Machine Learning, ICML}
  
  \vskip 0.3in
]



\printAffiliationsAndNotice{\icmlEqualContribution}
\input{sec/0_abstract}    
\input{sec/1_intro}
\input{sec/2_relatedworks}

\input{sec/3_method}
\input{sec/4_exp}
\input{sec/5_conclusion}
\section*{Impact Statement}

This paper presents work whose goal is to advance the field of machine learning. There are many potential societal consequences of our work, none of which we feel must be specifically highlighted here.
\bibliography{example_paper}
\bibliographystyle{icml2026}

%%%%%%%%%%%%%%%%%%%%%%%%%%%%%%%%%%%%%%%%%%%%%%%%%%%%%%%%%%%%%%%%%%%%%%%%%%%%%%%
%%%%%%%%%%%%%%%%%%%%%%%%%%%%%%%%%%%%%%%%%%%%%%%%%%%%%%%%%%%%%%%%%%%%%%%%%%%%%%%
% APPENDIX
%%%%%%%%%%%%%%%%%%%%%%%%%%%%%%%%%%%%%%%%%%%%%%%%%%%%%%%%%%%%%%%%%%%%%%%%%%%%%%%
%%%%%%%%%%%%%%%%%%%%%%%%%%%%%%%%%%%%%%%%%%%%%%%%%%%%%%%%%%%%%%%%%%%%%%%%%%%%%%%
\newpage
\appendix
\onecolumn
\input{sec/X_suppl}

%%%%%%%%%%%%%%%%%%%%%%%%%%%%%%%%%%%%%%%%%%%%%%%%%%%%%%%%%%%%%%%%%%%%%%%%%%%%%%%
%%%%%%%%%%%%%%%%%%%%%%%%%%%%%%%%%%%%%%%%%%%%%%%%%%%%%%%%%%%%%%%%%%%%%%%%%%%%%%%

\end{document}

%% file: sec/0_abstract.tex
\begin{abstract}
Current deep learning models for Multispectral and Hyperspectral Image Fusion (MS/HS fusion) are typically designed for fixed spectral bands and spatial scales, which limits their transferability across diverse sensors. To address this, we propose SSA, a universal framework for MS/HS fusion with spectral-band and fusion-scale agnosticism. Specifically, we introduce Matryoshka Kernel (MK), a novel operator that enables a single model to adapt to arbitrary numbers of spectral channels. Meanwhile, we build SSA upon an Implicit Neural Representation (INR) backbone that models the HS signal as a continuous function, enabling reconstruction at arbitrary spatial resolutions. Together, these two forms of agnosticism enable a single MS/HS fusion model that generalizes effectively to unseen sensors and spatial scales. Extensive experiments demonstrate that our single model achieves state-of-the-art performance while generalizing well to unseen sensors and scales, paving the way toward future HS foundation models.
\end{abstract}

%% file: sec/1_intro.tex
\section{Introduction}
\label{sec:intro}
Hyperspectral imaging (HSI) \cite{amigo2019hyperspectral} provides rich spectral information across hundreds of narrow bands, making it invaluable in domains like medical diagnostics \cite{fauvel2012advances}, remote sensing, and target detection \cite{uzair2013hyperspectral,technologies13050170}. However, a fundamental trade-off in optical systems means this rich spectral detail often comes at the cost of low spatial resolution, hampering downstream tasks such as classification \cite{he2017recent} and object tracking \cite{chen2021object}. To overcome this, Multispectral and Hyperspectral Image Fusion (MS/HS Fusion) has emerged as a key solution, aiming to reconstruct a high-resolution HSI by fusing a low-resolution HSI with a high-resolution multispectral image of the same scene.
In recent years, with the rapid development of deep learning, DL-based methods \cite{masi2016pansharpening,yang2017pannet,dian2018deep} have gradually become the mainstream for MS/HS Fusion. These methods can automatically learn complex nonlinear relationships, exhibiting immense potential in HSI fusion tasks.
\begin{figure}[!t]
    \centering
    \includegraphics[width=\linewidth]{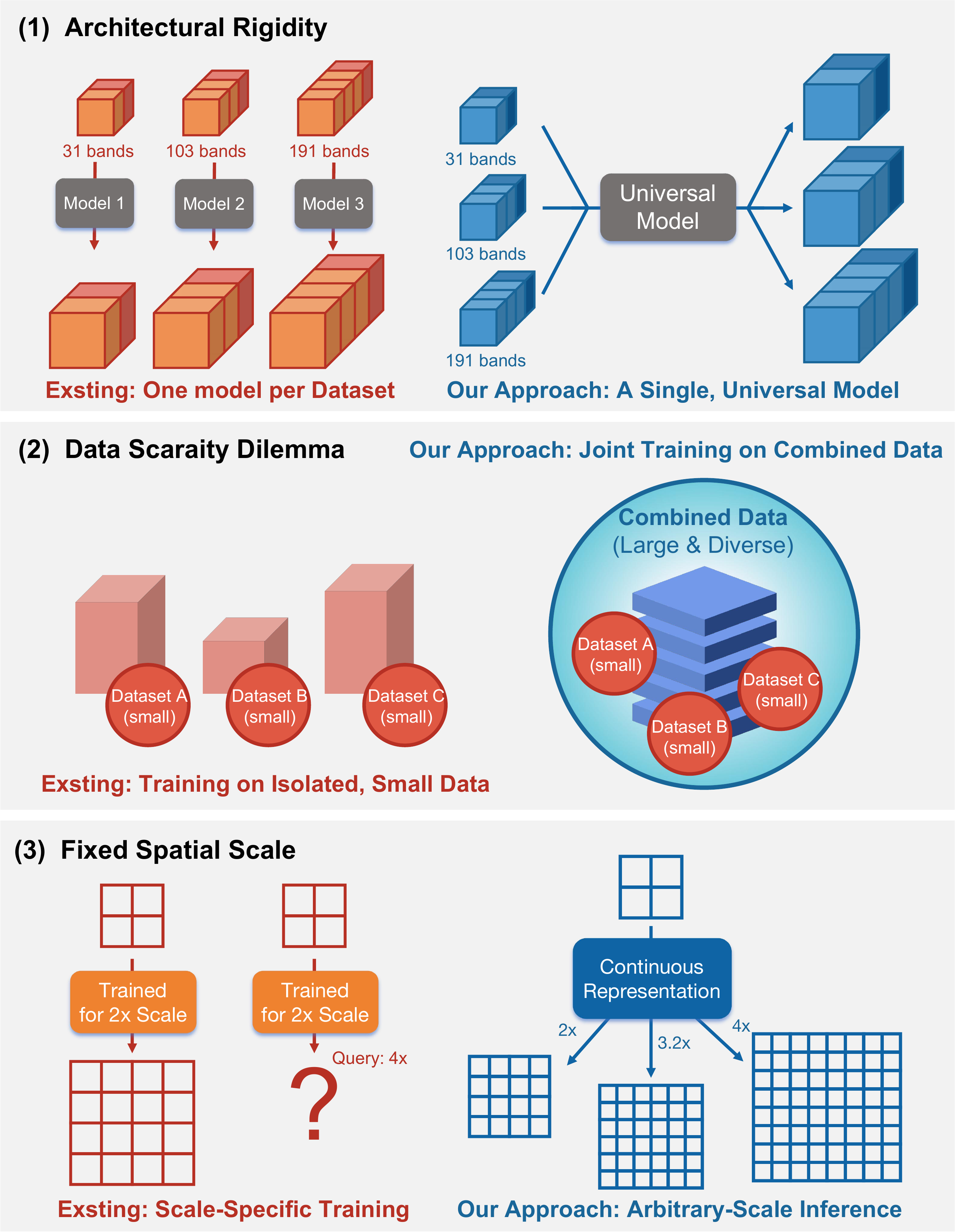}
    \caption{Existing paradigms \textit{v.s.} our universal approach for hyperspectral image fusion. Our method can suit different spectral bands joint training and can inference on arbitrary-scale fusion.}
    \label{fig:limitation}
    \vspace{-2em}
\end{figure}

Despite this significant progress, the real-world applicability of DL-based methods is hampered by the deep-rooted challenge of \textit{sensor diversity} \cite{hong2025hyperspectral}. Datasets captured by different sensors, such as AVIRIS (224 bands) or Pavia University (103 bands), vary drastically in their number of spectral bands, which exposes a fundamental lack of generality in current models. As illustrated in Fig.~\ref{fig:limitation}, this limitation manifests in three primary ways:
\textit{(1) Architectural Rigidity}: Most deep neural networks are designed with hard-coded channel dimensions, forcing researchers to either train independent models for each sensor~\cite{deng2023psrt, wu2025fully} or resort to costly fine-tuning~\cite{gonzalez2025multispectral}. Some even manually select band subsets~\cite{suryanarayana2025deep}, which sacrifices valuable spectral information.
\textit{(2) The Dilemma of Data Scarcity and Model Scale}: HSI datasets are typically small. Training on such isolated, small-scale data constrains model capacity and hinders the learning of generalizable features, leading to a high risk of overfitting~\cite{hoffmann2022training, yan2025hyperspectral}.
\textit{(3) Fixed Spatial Scaling Factor}: Current models~\cite{dengbidirectional, hu2022fusformer} are trained for a fixed integer upscaling factor (\textit{e.g.}, $4\times$, $8\times$). They learn a discrete grid-to-grid mapping, failing to handle the arbitrary or non-integer scales required in many applications.

To dismantle these barriers, we introduce SSA, a unified framework that addresses the above challenges through two technical innovations. To tackle spectral rigidity (1) and its impact on data scarcity (2), since hard-coded architectures prevent pooling data across sensors for large-scale training, we propose the Matryoshka Kernel (MK). Inspired by Matryoshka Representation Learning (MRL)~\cite{kusupati2022matryoshka}, MK allows a single model to process variable numbers of spectral bands, enabling joint training on heterogeneous datasets. Concurrently, to resolve spatial rigidity (3), we build our framework upon an Implicit Neural Representation (INR) backbone, which models signals as continuous functions and supports reconstruction at arbitrary (including non-integer) scaling factors.
In this work, we integrate these concepts into a cohesive MS/HS fusion model as a single unified framework. Our contributions are threefold:
\begin{itemize}
\item We introduce Matryoshka Kernels (MK), a novel architectural principle realized through adaptive convolution layers. This allows our model to natively process arbitrary numbers of spectral bands, enabling spectral-band agnosticism.
\item By building our framework upon an Implicit Neural Representation (INR), we replace discrete scaling with a continuous function, enabling reconstruction at arbitrary spatial scales.
\item Extensive experiments show that SSA achieves state-of-the-art performance while generalizing well to unseen sensors and scaling factors, substantially reducing the need for per-sensor retraining.
\end{itemize}

%% file: sec/2_relatedworks.tex
\section{Related Works}
\label{sec:related}

\subsection{Deep Learning for MS/HS Image Fusion}

Following PanNet~\cite{yang2017pannet}, which introduced deep learning to hyperspectral and multispectral (MS/HS) image fusion, the research paradigm has shifted from traditional methods, such as matrix factorization~\cite{kawakami2011high,huang2013spatial} and sparse coding~\cite{nezhad2016fusion}, to data-driven deep learning. Current research largely focuses on designing novel network architectures, for instance, by incorporating attention mechanisms and Transformers~\cite{mimo-sst,wu2025fully,cao2025efficient} or adopting popular paradigms like GANs~\cite{shang2024mft,zhou2022unsupervised} and diffusion models~\cite{Cao2024DDIF}. However, these highly specialized architectures are often deeply coupled with specific data specifications (e.g., the number of spectral bands), leading to a severe lack of universality and transferability.

% Recently, several works~\cite{zhang2022implicit,chen2023spectral} have begun to apply continuous signal representation (i.e., Implicit Neural Representation, INR) to HSI data. By directly mapping coordinates to a continuous function that models the hyperspectral cube, they achieve reconstruction at arbitrary scales, addressing the spatial rigidity of conventional models. Building on this, researchers have explored various avenues: CLORF~\cite{wang2025hyperspectral} leverages the low-rank nature of matrix factorization and the continuity of neural representations for self-supervised learning; FeINFN~\cite{liang2024fourier} performs implicit fusion in the frequency domain to enhance high-frequency details; OTIAS~\cite{deng2025otias} proposes an Octree-based implicit adaptive sampling to fuse spatial and spectral information; and AFO~\cite{zhu2025arbitrary} designs a novel integral as a mapping operator for a cross-arbitrary-scale fusion scheme. However, despite these pioneering INR-based methods solving the problem of spatial rigidity, they still need to re-engineer the model for datasets with different numbers of bands. Moreover, as INR is a typical data-driven method, it may not be easy to train a model with better generalization given the current data volume.

\begin{figure*}[!ht]
    \centering
    \includegraphics[width=\textwidth]{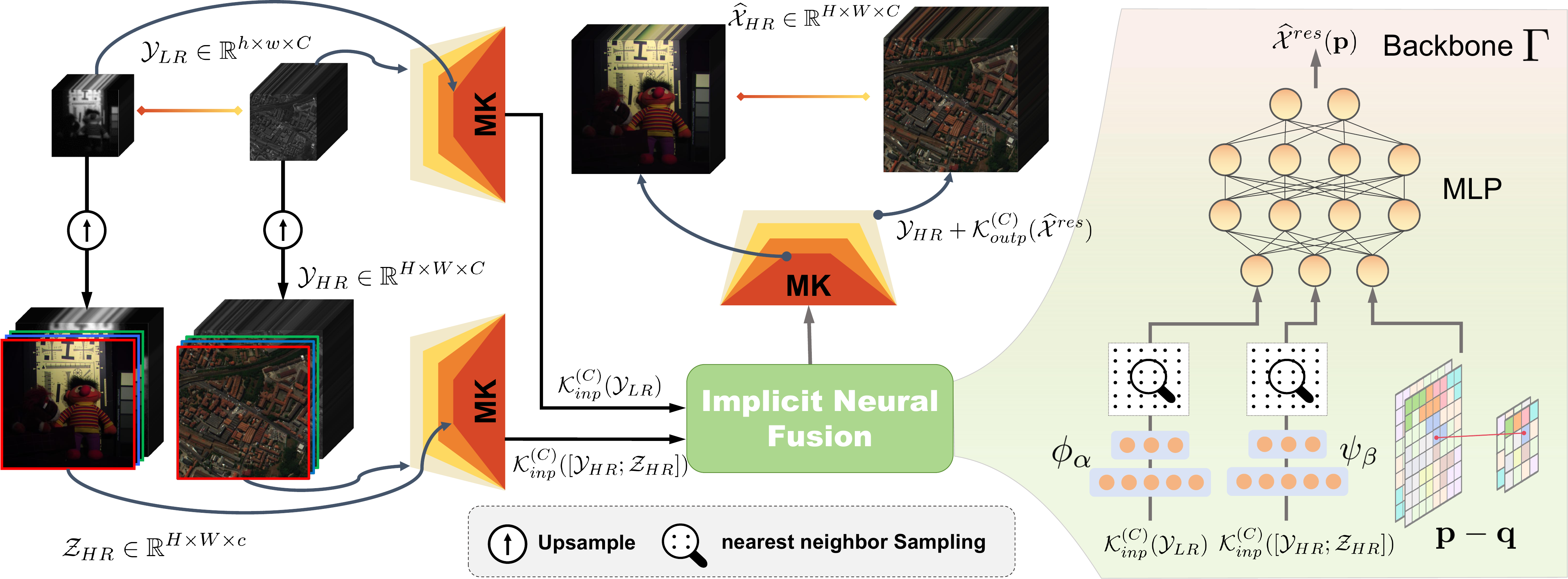}
    \caption{The overall architecture of our proposed SSA framework. This end-to-end model realizes the universal mapping function $\mathcal{F}$, which takes an LR-HSI and an HR-MSI from \textit{various sensors} as input and reconstructs a high-fidelity HR-HSI at \textit{arbitrary scales}.}
    \label{fig: architecture}
\end{figure*}

Recently, several works~\cite{zhang2022implicit,chen2023spectral} have begun to apply Implicit Neural Representation (INR) to HSI data, modeling the image as a continuous function of coordinates. This approach enables reconstruction at arbitrary scales, neatly addressing the spatial rigidity of conventional models. Subsequent research has explored various extensions, such as integrating low-rank properties~\cite{wang2025hyperspectral}, performing fusion in the frequency domain~\cite{liang2024fourier}, or using adaptive sampling~\cite{deng2025otias}. However, despite solving spatial rigidity, these methods face two key limitations: First, they lack spectral flexibility, as models must be re-engineered for datasets with different numbers of bands. Second, as data-driven approaches, achieving strong generalization can be difficult given the limited volume of available HSI data.

\subsection{Matryoshka Representation Learning and Flexible Model Design}
Our solution is philosophically inspired by the broader trend in machine learning towards building more adaptive and general-purpose models. From the perspective of representation granularity adaptation, Matryoshka Representation Learning (MRL)~\cite{kusupati2022matryoshka,hojjat2025thinkingvit,cappellazzo2025mome} was proposed to explore model flexibility. The core idea of MRL is to train a single, high-dimensional feature vector $z \in \mathbb R^d$ that encodes information at different granularities in a nested fashion, such that its shorter prefixes also serve as valid representations. This paradigm provides flexible representations for downstream tasks with varying computational demands.

Formally, a nested representation $z$ obtained by a deep neural network $z:=F(\cdot,\theta_F): \mathcal X\to \mathbb R^d$ contains different granularities representations: $z_{1:m}\in \mathbb R^m$, where $m$ is the desired dimension and $(r:s)$ means one sub-representation choosing algorithm. The Matryoshka representation is acquired to minimize the empirical risk (that is, a valid representation). Under the classification context and parameterizing the nested $z$ as the classifier weights $\mathbf W^{(m)} \in \mathbb R^{N, m}$, the expirical risk can be supervised using cross-entropy loss:
\begin{equation}
    \min_{\substack{\theta_F, \\ \{\mathbf W^{(m)}\}_{\substack{m\in \mathcal M, \\ m\leq d}}}}\frac{1}{N} \sum_{m \in \mathcal M} \mathcal L\left(\mathbf W^{(m)} \cdot F(x,\theta_F)_{1:m}; y \right), \label{eq:mrl-loss}
\end{equation}
where $\mathcal M$ is a nested chosen list (\textit{e.g,} $\mathcal M:=\{m_1,\dots,m_i,d\}$, $m_1\leq m_i\leq d$), $N$ is the total classification number, $(x,y)$ is the data/label pair. In the MRL paper, the sub-representation choosing algorithm is designed as dim-slicing, where $r$ is fixed to 1 and $s=m$.

% enabling adaptive deployment per computational constraints.
% This representation strategy has recently received widespread attention, with many works leveraging shared models to construct embeddings of different granularities. For example, ThinkingViT~\cite{hojjat2025thinkingvit} integrates a nested structure to dynamically adjust its computational budget based on input image complexity. MoME~\cite{cappellazzo2025mome} combines MRL with Mixture-of-Experts for audio-visual speech recognition. MatryoshkaKV~\cite{lin2024matryoshkakv} introduces an adaptive method that compresses the key-value cache along the feature dimension in large language models through trainable orthogonal projections and a Matryoshka-style training strategy. fMRLRec~\cite{wang2024train} utilizes linear recurrent units and a full-scale Matryoshka operator to fuse text and image features, learning item representations of multiple granularities in a single training run.

% The principle of achieving flexibility through a nested, structured representation, as demonstrated by MRL, provides powerful inspiration for our work. 
Different from the MRL, which aims at learning flexible valid representation, and inference with different size model dimension~\cite{lin2024matryoshkakv,wang2024train}, we bring the nested principle into the hyperspectral data fusion kernels to handle various input image channels, termed as Matryoshka Kernels (MK, see \S  \ref{subsec:mk}).
This convertion eliminates  previous two key drawbacks: architectural rigidity, data scarity, and model scale.  

% We term this architectural innovation Matryoshka Kernels (MK). Unlike MRL, which creates nested embeddings, our spectral-agnostic layers create nested convolutional kernels. This allows the model architecture itself to dynamically and non-parametrically adapt to a fundamental property of the input data—namely, its number of spectral channels. This architectural adaptability offers a direct solution to the problem of input structural heterogeneity.

% In summary, SA$^2$ is the first framework to synergize the established spatial flexibility of the INR paradigm with the novel spectral flexibility afforded by our proposed Matryoshka Kernels. By doing so, it overcomes a core limitation of prior HSI-SR approaches and lays the groundwork for the development of a general-purpose foundation model for hyperspectral image analysis.

%% file: sec/3_method.tex
\section{Method}
\label{sec:method}
\subsection{Problem Formulation and Model Overview}

\newcommand{\xhr}{\mathcal{X}_{HR}}
\newcommand{\wxhr}{\widehat{\mathcal{X}}_{HR}}
\newcommand{\ylr}{\mathcal{Y}_{LR}}
\newcommand{\yhr}{\mathcal Y_{HR}}
\newcommand{\zhr}{\mathcal{Z}_{HR}}
\newcommand{\deltax}{\Delta \widehat{\mathcal X}}
\newcommand{\deltaxp}{\Delta \widehat{\mathcal X}^\prime}
\newcommand{\kinp}{\mathcal K_{inp}}
\newcommand{\koup}{\mathcal K_{outp}}

Given a low-resolution HSI (LR-HSI), $\mathcal{Y}_{LR} \in \mathbb{R}^{h \times w \times C}$, and a high-resolution multispectral image (HR-MSI), $\mathcal{Z}_{HR} \in \mathbb{R}^{H \times W \times c}$ ($c \ll C$), the objective of MS/HS fusion is to learn a mapping function $\mathcal{F}_\theta: (\ylr, \zhr)\to\wxhr$ that reconstructs a high-resolution hyperspectral image (HR-HSI), $\mathcal{X}_{HR} \in \mathbb{R}^{H \times W \times C}$, where $\theta$ represents the learnable parameters of the model.

A key challenge in building a universal fusion model is sensor heterogeneity. For different sensors and optical cameras, the number of spectral bands $C$ and the spatial scaling factor $s$ can vary. This means our training data comes from a collection of datasets $\mathcal D=\{D_1, \cdots, D_N\}$, and for any image pair $(\ylr^{(i)}, \zhr^{(i)})$ sampled from the $i$-th dataset, it has a corresponding number of spectral bands $C^{(i)}$ and a scaling factor $r^{(i)}$. This requires the universal fusion model should acquire the spectral and spatial agnosticism.

To realize such a universal $\mathcal F$, we propose the SSA framework, with its overall architecture depicted in Fig.~\ref{fig: architecture}. Our model is designed as an end-to-end, continuous representation-based fusion pipeline.
First, the model performs spectral unification given data pair.
As shown, any input $\ylr$, regardless of its original band count $C$, is mapped by a Matryoshka Kernel Layer (MKL) $\mathcal K_{inp}^{(C)}(\cdot)$ to a fixed, predefined feature dimension (see \S \ref{subsec:mk}).
Concurrently, the HR-MSI $\zhr$ is concatenated with the spatial upsampled LR-HSI $\yhr\in \mathbb R^{H\times W\times c}$, and this combined tensor is processed by a similar layer.

Subsequently, those two feature maps are fed into the continuous representation fusion backbone (see \S \ref{subsec:inr}). The core of this module is to model an implicit function, parameterized by an MLP, which is responsible for the deep fusion of spectral and spatial information.
The continuous feature output by the fusion backbone is then fed to an output MKL $\koup^{(C)}(\cdot)$ to the pixel space with original band count $C$. 
Finally, inspired by residual learning in \cite{he2016deep}, the output $\mathcal{\widehat X}^{res}$ is added with the upsampled LR-HSI to produce the final high-fidelity HR-HSI.

\subsection{Matryoshka Kernels: Spectral Agnosticism}
\label{subsec:mk}
\begin{figure}[!h]
    \centering
    \includegraphics[width=\linewidth]{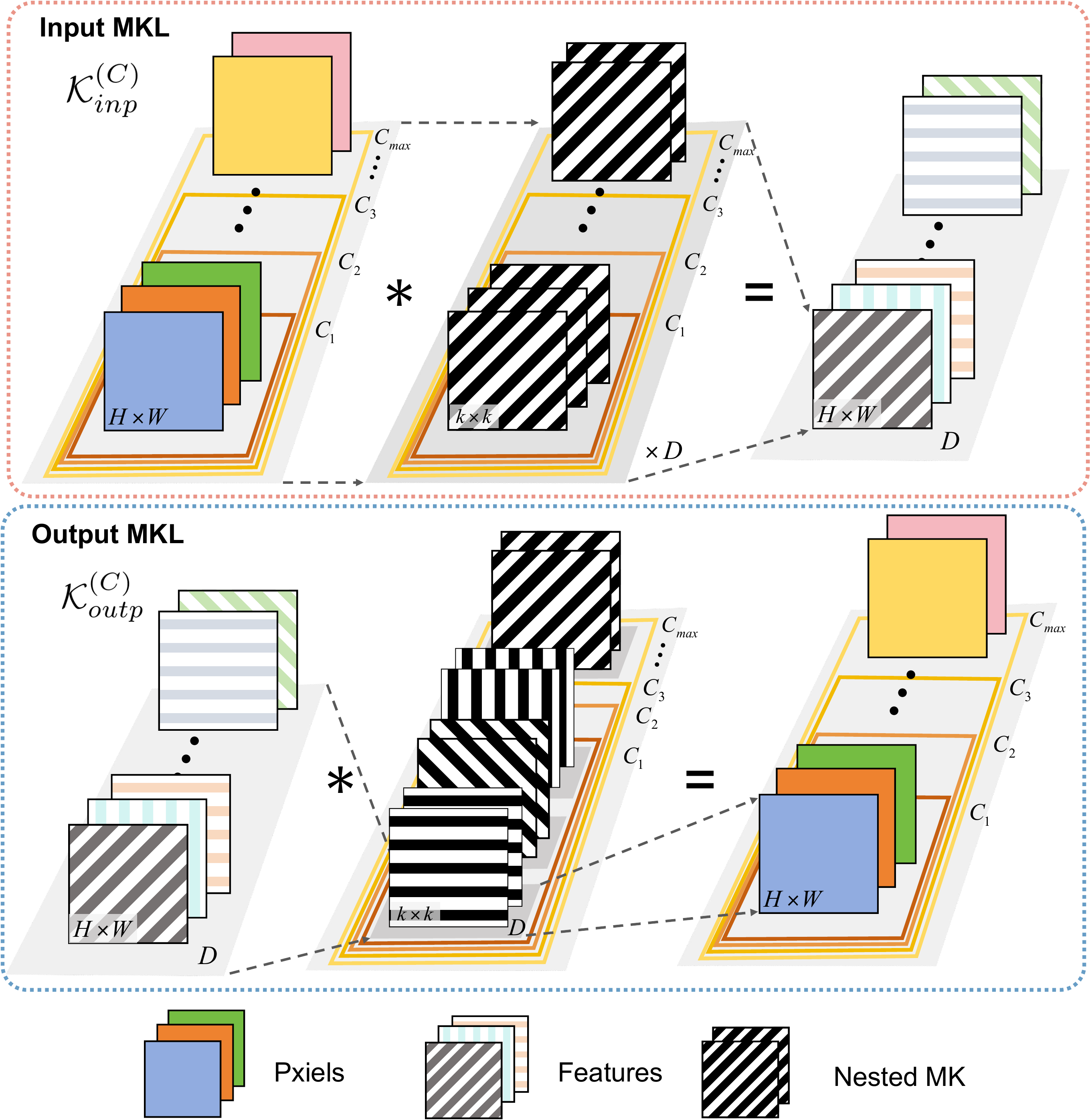}
    \caption{Illustration of the MKL's encoding and decoding.}
    \label{fig:placeholder}
\end{figure}

To deal with the drawbacks mentioned \S~\ref{sec:intro}, particularly when dealing with inputs of varying channels, we introduce the Matryoshka Kernel Layer (MKL) formulated as $\mathcal K_{\star}^{(C)},\star\in\{inp, outp\}$, which acts as input and output layers. Each MKL contains one Matryoshka-style nested kernel, see Fig.~\ref{fig:placeholder}.
This approach avoids the need for separate layers for each possible channel configuration. 
For the input MKL, given a hyperspectral cube $\mathbf{X}$ with $C$ bands, it aims to embed the input into the feature $\mathbf Y$ with $D$-embedding dimension. For the output MKL, it's vice versa.

Specifically, each MKL internally maintains an MK as the learnable weight. Taking the input MKL weight as example,  the MK's weight is as $\mathbf{W}_{\text{nested}} \in \mathbb{R}^{D \times C_{\text{max}} \times k \times k}$, where $D$ is the fixed number of output channels, $C_{\text{max}}$ is the predefined maximum number of input channels the layer can support ($C_{\text{in}} \le C_{\text{max}}$), and $k \times k$ is the kernel size. A corresponding bias vector $\mathbf{b} \in \mathbb{R}^{D}$ is also maintained.
During the forward pass, a valid kernel is generated from the nested kernel, $\mathbf{W}_{\text{valid}} \in \mathbb{R}^{D \times C_{\text{in}} \times k \times k}$, using a slice-based approach:
\begin{equation}
    \mathbf{W}_{\text{valid}} = \mathbf{W}_{\text{nested}}[:, :C_{\text{in}}, :, :],
    \label{eq:input_slice}
\end{equation}
where operation $[\,:\,]$ follows Numpy array slice, which selects the first $C_{\text{in}}$ items along the input channel dimension of $\mathbf{W}_{\text{nested}}$ to form the kernel for the current computation.
Using the current valid kernel, the final output feature map can be produced by applying a standard 2D convolution:
\begin{equation}
    \mathbf{Y} = \texttt{Conv2D}(\mathbf{X}, \mathbf{W}_{\text{valid}}, \mathbf{b}, \text{stride}, \text{padding}).
\end{equation}
For each element in the $d$-th output channel, the computation is formally expressed as:
\begin{equation}
    \mathbf{Y}_{d, i, j} = \mathbf{b}_d + \sum_{c=0}^{C_{\text{in}}-1} \sum_{m=0}^{k-1} \sum_{n=0}^{k-1} \mathbf{X}_{c, i \cdot s+m, j \cdot s+n} \cdot (\mathbf{W}_{\text{valid}})_{d, c, m, n}, \notag
\end{equation}
where $s$ denotes the stride and the indices $d$, $c$, $m$, and $n$ refer to the output channel, input channel, and spatial dimensions of the kernel, respectively.
% The key to our model's ability to handle a varying number of spectral bands lies in our proposed concept of Matryoshka Kernels (MK). This concept is realized through our custom Spectral-Agnostic Layers, the core idea of which is to learn a ``superset'' kernel.

A similar nested philosophy is applied to the output MKL of the model, enabling it to dynamically generate an output that matches the original band count $C$ of the input sample.
During the forward pass, a desired output channel count, $C_{\text{out}}$ ($C_{\text{out}} \le C_{\text{max}}$), is specified. The valid kernel $\mathbf{W}_{\text{valid}} \in \mathbb{R}^{C_{\text{out}} \times D \times k \times k}$ and bias $\mathbf{b}_{\text{comp}} \in \mathbb{R}^{C_{\text{out}}}$ are generated by slicing the superset parameters along the output channel dimension:
\begin{align}
    \mathbf{W}_{\text{valid}} &= \mathbf{W}_{\text{nested}}[:C_{\text{out}}, :, :, :], \\
    \mathbf{b}_{\text{valid}} &= \mathbf{b}_{\text{nested}}[:C_{\text{out}}].
    \label{eq:output_slice_w}
\end{align}
Through this mechanism of a nested kernel and dynamic subset slicing, our SSA model achieves true spectral agnosticism, allowing it to seamlessly process heterogeneous data from any sensor within a universal architecture. The detailed input and output MKL algorithms are shown in App.\S~\ref{mkl_algos}.
\input{Tabs/tab_cave_paviac}
\subsection{Arbitrary-Scale Image Fusion as Continuous Function}
\label{subsec:inr}
The implementation of our target HR-HSI as a continuous function through the implicit fusion backbone $\Gamma$ is what grants our model true spatial flexibility. It allows the model to be queried on any coordinate grid, enabling the generation of an output at any desired resolution.

As illustrated in Fig.~\ref{fig: architecture} (b), our model first maps the embedded features after MK into a latent space via two parallel encoders $ \phi_\alpha,  \psi_\beta$. This process can be represented as:
\begin{equation}
\begin{cases} 
\mathcal E_{pe} = \phi_{\alpha}(\mathcal K_{inp}^{(C)}(\ylr)), \\
\mathcal{E}_{pa} = \psi_{\beta}\left(\mathcal K_{inp}^{(C)}([\mathcal{Y}_{HR}; \mathcal{Z}_{HR}])\right), \notag
\end{cases}
\end{equation}
where $[\cdot;\cdot]$ stands for channel concatenation. These latent codes, $\mathcal{E}_{pe}$ and $\mathcal{E}_{pa}$, provide both the spectral and spatial semantic information necessary for the subsequent continuous function modeling.

To predict the spectral vector for any continuous coordinate $\mathbf{p}$ in the target HR-HSI space, our fusion module performs a local, query-based process. For a given pixel, we use its center to represent its coordinate and map the HR coordinates to a normalized square grid $\Omega = [-1,1]^2$. The integer coordinates $(i,j)$ of a pixel in the $H\times W$ source grid are then mapped to a query coordinate $\mathbf p_{(i,j)}\in \Omega \subset \mathbb R^2$ using the following transformation:
\begin{equation}
    \mathbf{p}_{(i,j)} = \left( \frac{2i}{H-1} - 1, \frac{2j}{W-1} - 1 \right).
\end{equation}
Using the query coordinate $\mathbf{p}$ as a pivot, we sample a precise local spatial feature from the spatial latent code $\mathcal{E}_{pa}$. Concurrently, we identify the nearest neighbor $\mathbf{q}_{(i,j)}\in \Omega$ corresponding to $\mathbf{p}_{(i,j)}$ in the low-resolution spectral latent code $\mathcal{E}_{pe}$ and sample the features of this neighbor.

For the sampled local spectral features, we feed the previous spectral, spatial codes and the relative coordinates $\mathbf{p}-\mathbf{q}$ that describe the fine-grained position into a shared MLP decoder:
\begin{equation}
    \widehat{\mathcal{X}}^{res}(\mathbf{p}) = \texttt{MLP}(\mathcal{E}_{pe}(\mathbf{p}), \mathcal{E}_{pa}(\mathbf{q}), \mathbf{p}-\mathbf{q}),
\end{equation}
where $\mathbf{p}$ is the normalized coordinates of each query pixel in the HR domain. The output of the MLP is then fed into the output MK layer. Leveraging the MK, this layer dynamically maps the MLP feature back to the original spectral band count $C$ of the current sample, generating the final predicted residual components $\mathcal K_{outp}^{(C)}(\widehat{\mathcal{X}}^{res}(\mathbf{p}))$. The complete fused image is finally obtained by adding the predicted residual to the bicubic-interpolated base image:
\begin{equation}
    \widehat{\mathcal{X}}_{HR}(\mathbf{p}) = \mathcal{Y}_{HR}(\mathbf{p}) + \mathcal K_{outp}^{(C)}(\widehat{\mathcal{X}}^{res}(\mathbf{p})).
\end{equation}

Additionally, to ensure spatial continuity when crossing grid boundaries, we follow the practice of LIIF~\cite{chen2021learning}. We identify the four-nearest neighbor grid points $\mathbf q_{i}, i\in \{1,\dots 4\}
$ in the low-resolution feature map. During decoding, we predict a candidate spectral residual $\widehat{\mathcal{X}}^{res}_i(\mathbf{p})$ and a scalar weight $w_i$ for each neighbor $q_i$. The final residual vector is then obtained by a softmax weighted fusion of all candidate residuals:
\begin{equation}
    \widehat{\mathcal{X}}^{res}(\mathbf{p}) = \sum_{i=1}^4 \frac{\exp(w_i)}{\sum_{j=1}^4 \exp(w_j)} \widehat{\mathcal{X}}^{res}_i(\mathbf{p}).
\end{equation}
True scale agnosticism is achieved through this coordinate-based approach. By modeling the HR-HSI as a continuous function, our architecture avoids fixed upsampling layers, granting it the ability to generate outputs at any resolution.
\begin{figure*}[!ht]
    \centering
    \includegraphics[width=\textwidth]{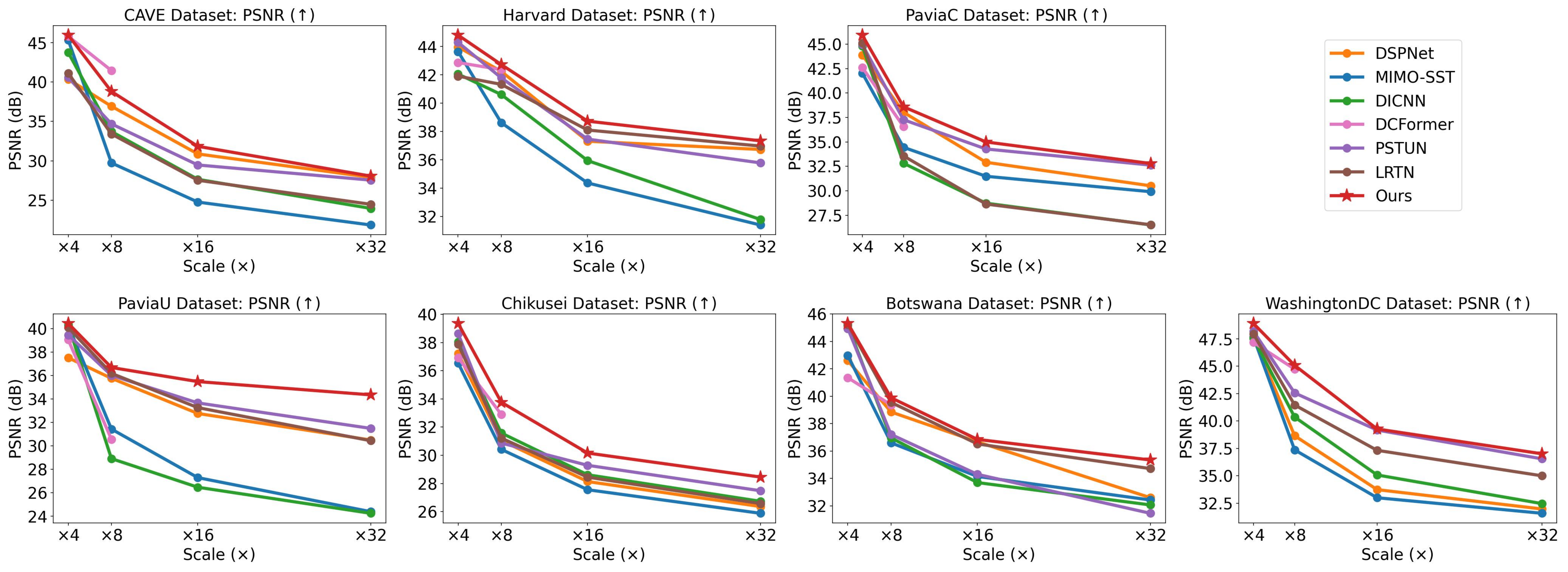}
    \caption{Quantitative comparison of PSNR (dB) across multiple scaling factors on all seven datasets.}
    \label{fig:psnr}
\end{figure*}
\subsection{Cross-Sensor Joint Training Strategy}
\label{subsec:training}

The conventional ``one dataset, one model'' paradigm is not only inefficient but also prone to being easily overfitting~\cite{yan2025hyperspectral}, especially given the small scale of HSI datasets. Training a universal model within various hyperspectral data usually faces a challenge that different bands and spatial sizes can not be batched together to enable joint training.

Our training pipeline is designed to handle $N$ heterogeneous datasets, denoted as $\mathcal{D} = \{D_1, \dots, D_N\}$.
We first partition all training samples into distinct buckets, where each bucket  $D_i$ contains data from a single source dataset, thus sharing identical spatial dimensions and number of spectral bands. 
At each training step, we construct a mini-batch by a two-stage sampling process. First, a bucket $D_i$ is selected from $\mathcal D$ following a uniform random distribution. Then, a batch of samples is drawn exclusively from the chosen bucket. This strategy ensures that all samples within a single mini-batch are structurally homogeneous, while samples across different mini-batches can be heterogeneous.
It is precisely this batch-to-batch heterogeneity that our model is designed for.
% Thanks to our Matryoshka Kernels (\S\ref{subsec:mk}) and Continuous Representation (\S\ref{subsec:inr}), the model seamlessly adapts to the structural differences from one batch to the next. 
Crucially, all samples are processed by a universal model parameters $\theta$, which are optimized via a unified hybrid reconstruction loss function.
Following Eq.~\eqref{eq:mrl-loss}, the reconstruction loss can be written in the MRL style:
\begin{equation}
    \min_{\substack{\theta,\, C\in \text{Dim}(D_i), \\ (\ylr,\zhr)\in D_i}} \frac 1 N \sum_{i=1}^N  \mathcal L_{total}\left[ \mathcal F_\theta( \ylr,\zhr); \xhr \right],  \notag 
\end{equation}
where $\mathcal F_\theta$ is the model (\textit{i.e.}, $\mathcal F_\theta:=\mathcal K_{outp}^{(C)} \circ \Gamma \circ \mathcal K_{inp}^{(C)}$ and $\circ$ means the network composition), Dim$(\cdot)$  denotes the bands numbers in the data bucket. Our total loss, $\mathcal{L}_{total}$, is a weighted sum of the L1 loss and the Structural Similarity (SSIM) loss, defined as:
\begin{equation}
\begin{aligned}
    &\mathcal{L}_{total} = \mathbb{E}_{(\mathcal{Y}_{LR}, \mathcal{Z}_{HR}, \mathcal{X}_{HR}) \sim D_{i}} \left[ \mathcal{L}_{1}+ \lambda\mathcal{L}_{SSIM} \right], \\
    &\mathcal{L}_{1} = \| \mathcal{F}_\theta(\mathcal{Y}_{LR}, \mathcal{Z}_{HR})  - \mathcal{X}_{HR} \|_1 ,\\
    &\mathcal{L}_{SSIM} = 1 - \text{SSIM}(\mathcal{F}_\theta(\mathcal{Y}_{LR}, \mathcal{Z}_{HR}), \mathcal{X}_{HR}),
\end{aligned}
\end{equation}
where $\lambda$ is a weighting coefficient, $\| \cdot \|$ is the $\ell_1$ norm, and $\text{SSIM}(\cdot, \cdot)$ is the structural similarity index measure.

Through this joint training strategy, our SSA model is forced to learn a universal fusion mapping rather than memorizing the specifics of a particular dataset.
This approach endows the model with remarkable generalization capabilities to unseen sensors (see \S~\ref{sec:generalization}).

%% file: Tabs/tab_cave_paviac.tex
% % Define custom colors for highlighting
% \definecolor{bestcolor}{RGB}{255, 224, 224}
% \definecolor{secondcolor}{RGB}{224, 224, 255}
% \definecolor{normalcolor}{RGB}{255, 255, 255} % White for normal cells

% % --- NEW COMMANDS FOR SINGLE-CELL RENDERING ---
% % This command creates a perfectly aligned value ± std pair in a single cell
% % It uses an inner tabular to achieve alignment on the ± symbol
% \newcommand{\perfalign}[3]{% #1: value, #2: symbol, #3: std
%     \begin{tabular}{@{}r@{\,}l@{}}
%         #1 & #2\ #3
%     \end{tabular}
% }
% \newcommand{\perfalignplain}[2]{% #1: value, #2: std (for DCFormer)
%     \begin{tabular}{@{}r@{\ }l@{}}
%         #1 & #2
%     \end{tabular}
% }

% % Commands to apply background color to a SINGLE CELL
% \newcommand{\best}[1]{\cellcolor{bestcolor}\perfalign{#1}}
% \renewcommand{\second}[1]{\cellcolor{secondcolor}\perfalign{#1}}
% \newcommand{\normal}[1]{\cellcolor{normalcolor}\perfalign{#1}}
% \newcommand{\bestplain}[1]{\cellcolor{bestcolor}\perfalignplain{#1}}
% \newcommand{\secondplain}[1]{\cellcolor{secondcolor}\perfalignplain{#1}}

\begin{table*}[!ht]
\centering
\renewcommand{\arraystretch}{1.0}
\setlength{\tabcolsep}{1pt}
\caption{The average and standard deviation calculated for all the compared approaches on 10 CAVE examples and 2 the PaviaC dataset examples simulating $\times4$ (in-distribution) and $\times 8,\times 16,\times 32$ (\textit{out-of-distribution}). Results are highlighted as 
    \colorbox{bestcolor}{best} and \colorbox{secondcolor}{second best}.}
\label{tab:cave_pavia_quantitative_comparison}
% \sisetup{}

\resizebox{\textwidth}{!}{%
% Each model now uses a single centered 'c' column
\begin{tabular}{@{}l l l 
  c c c c c c c | c c c c c c c @{}}
\toprule
\midrule
\multicolumn{3}{c}{\multirow{2}{*}{\textbf{Metric}}} & \multicolumn{7}{c}{\textbf{CAVE Dataset}} & \multicolumn{7}{c}{\textbf{PaviaC Dataset}} \\

\cmidrule(lr){4-10} \cmidrule(l){11-17}
\multicolumn{3}{c}{} 
& \textbf{DSPNet} & \textbf{MIMO-SST} & \textbf{DCINN} & \textbf{DCFormer} & \textbf{PSTUN} & \textbf{LRTN} & \textbf{Ours}
& \textbf{DSPNet} & \textbf{MIMO-SST} & \textbf{DCINN} & \textbf{DCFormer} & \textbf{PSTUN} & \textbf{LRTN} & \textbf{Ours} \\
\midrule
% --- 4× In-Dist. 数据行 ---
\multirow{4}{*}{\rotatebox[origin=c]{90}{In-Dist.}} & \multirow{4}{*}{$\times$4}
& PSNR($\uparrow$) 
& \normal{40.37}{$\pm$}{3.25} & \normal{45.35}{$\pm$}{3.35} & \normal{43.75}{$\pm$}{3.77} & \secondplain{45.81}{$\pm$}{5.56} & \normal{40.57}{$\pm$}{3.72} & \normal{41.13}{$\pm$}{3.24} & \best{45.96}{$\pm$}{4.69}
& \normal{43.89}{$\pm$}{1.78} & \normal{42.04}{$\pm$}{0.78} & \normal{44.79}{$\pm$}{1.99} & \normal{42.61}{$\pm$}{1.19} & \normal{44.94}{$\pm$}{1.50} & \secondd{45.13}{$\pm$}{1.01} & \best{45.92}{$\pm$}{1.75} \\
& & SAM($\downarrow$) 
& \normal{2.72}{$\pm$}{0.89} & \normal{2.74}{$\pm$}{0.86} & \normal{2.72}{$\pm$}{0.81} & \secondplain{2.70}{$\pm$}{1.03} & \normal{2.94}{$\pm$}{0.89} & \normal{2.92}{$\pm$}{1.38} & \best{2.55}{$\pm$}{1.00}
& \normal{3.15}{$\pm$}{0.43} & \normal{3.29}{$\pm$}{0.45} & \secondd{2.59}{$\pm$}{0.41} & \normal{3.55}{$\pm$}{0.43} & \normal{2.95}{$\pm$}{0.46} & \normal{3.18}{$\pm$}{0.45} & \best{2.02}{$\pm$}{0.30} \\
& & ERGAS($\downarrow$) 
& \normal{2.51}{$\pm$}{0.71} & \normal{2.08}{$\pm$}{0.68} & \normal{2.37}{$\pm$}{0.95} & \secondplain{2.01}{$\pm$}{1.82} & \normal{2.70}{$\pm$}{0.89} & \normal{2.38}{$\pm$}{0.82} & \best{1.72}{$\pm$}{1.13}
& \normal{1.58}{$\pm$}{0.10} & \normal{1.73}{$\pm$}{0.02} & \normal{1.34}{$\pm$}{0.12} & \normal{1.82}{$\pm$}{0.05} & \normal{1.47}{$\pm$}{0.07} & \normal{1.48}{$\pm$}{0.01} & \best{1.07}{$\pm$}{0.11} \\
& & SSIM($\uparrow$) 
& \normal{0.993}{$\pm$}{0.005} & \best{0.994}{$\pm$}{0.005} & \normal{0.992}{$\pm$}{0.004} & \secondplain{0.993}{$\pm$}{0.004} & \normal{0.992}{$\pm$}{0.005} & \normal{0.992}{$\pm$}{0.006} & \best{0.994}{$\pm$}{0.005}
& \normal{0.988}{$\pm$}{0.000} & \normal{0.986}{$\pm$}{0.000} & \secondd{0.992}{$\pm$}{0.000} & \normal{0.983}{$\pm$}{0.000} & \normal{0.990}{$\pm$}{0.000} & \normal{0.988}{$\pm$}{0.000} & \best{0.996}{$\pm$}{0.000} \\
\cmidrule(lr){1-17}
% --- 8× Out-of-Dist. 数据行 ---
\multirow{12}{*}{\rotatebox[origin=c]{90}{Out-of-Dist.}} & \multirow{4}{*}{$\times$8}
& PSNR($\uparrow$) 
& \normal{36.92}{$\pm$}{3.47} & \normal{29.73}{$\pm$}{3.21} & \normal{33.72}{$\pm$}{3.39} & \bestplain{41.46}{$\pm$}{4.98} & \normal{34.69}{$\pm$}{3.17} & \normal{33.40}{$\pm$}{3.13} & \secondd{38.80}{$\pm$}{3.79}
& \normal{38.03}{$\pm$}{0.80} & \normal{34.45}{$\pm$}{0.55} & \normal{32.82}{$\pm$}{0.58} & \normal{36.57}{$\pm$}{0.88} & \secondd{37.27}{$\pm$}{0.93} & \normal{33.57}{$\pm$}{0.40} & \best{38.59}{$\pm$}{0.67} \\
& & SAM($\downarrow$) 
& \normal{3.82}{$\pm$}{1.59} & \normal{5.55}{$\pm$}{2.66} & \normal{4.33}{$\pm$}{1.94} & \bestplain{3.65}{$\pm$}{1.65} & \normal{4.30}{$\pm$}{1.59} & \normal{4.55}{$\pm$}{2.03} & \secondd{3.70}{$\pm$}{1.45}
& \normal{5.66}{$\pm$}{0.25} & \normal{6.70}{$\pm$}{0.14} & \normal{6.09}{$\pm$}{0.19} & \secondd{5.49}{$\pm$}{0.26} & \normal{5.48}{$\pm$}{0.20} & \normal{6.67}{$\pm$}{0.15} & \best{4.74}{$\pm$}{0.06} \\
& & ERGAS($\downarrow$) 
& \normal{3.91}{$\pm$}{1.44} & \normal{9.91}{$\pm$}{4.03} & \normal{6.24}{$\pm$}{2.62} & \bestplain{3.08}{$\pm$}{2.21} & \normal{5.13}{$\pm$}{1.93} & \normal{6.00}{$\pm$}{2.03} & \secondd{3.46}{$\pm$}{1.57}
& \normal{4.05}{$\pm$}{0.05} & \normal{4.53}{$\pm$}{0.07} & \normal{4.53}{$\pm$}{0.09} & \normal{3.79}{$\pm$}{0.06} & \secondd{3.60}{$\pm$}{0.06} & \normal{4.51}{$\pm$}{0.05} & \best{3.47}{$\pm$}{0.05} \\
& & SSIM($\uparrow$) 
& \normal{0.980}{$\pm$}{0.015} & \normal{0.904}{$\pm$}{0.099} & \normal{0.958}{$\pm$}{0.029} & \bestplain{0.985}{$\pm$}{0.014} & \normal{0.964}{$\pm$}{0.022} & \normal{0.956}{$\pm$}{0.030} & \secondd{0.982}{$\pm$}{0.013}
& \normal{0.947}{$\pm$}{0.004} & \normal{0.923}{$\pm$}{0.006} & \normal{0.929}{$\pm$}{0.005} & \normal{0.948}{$\pm$}{0.004} & \normal{0.953}{$\pm$}{0.003} & \normal{0.931}{$\pm$}{0.006} & \best{0.960}{$\pm$}{0.004} \\
\cmidrule(lr){2-17}
% --- 16× Out-of-Dist. 数据行 ---
& \multirow{4}{*}{$\times$16}
& PSNR($\uparrow$) 
& \secondd{30.87}{$\pm$}{3.46} & \normal{24.76}{$\pm$}{2.93} & \normal{27.65}{$\pm$}{3.43} & - & \normal{29.47}{$\pm$}{3.16} & \normal{27.53}{$\pm$}{2.97} & \best{31.84}{$\pm$}{3.47}
& \normal{32.92}{$\pm$}{0.43} & \normal{31.49}{$\pm$}{0.20} & \normal{28.74}{$\pm$}{0.05} & - & \secondd{34.28}{$\pm$}{0.39} & \normal{28.65}{$\pm$}{0.11} & \best{35.00}{$\pm$}{0.02} \\
& & SAM($\downarrow$) 
& \secondd{6.51}{$\pm$}{2.83} & \normal{8.36}{$\pm$}{4.48} & \normal{6.93}{$\pm$}{3.48} & - & \normal{6.68}{$\pm$}{2.44} & \normal{6.73}{$\pm$}{3.08} & \best{5.93}{$\pm$}{2.42}
& \normal{8.43}{$\pm$}{0.39} & \normal{8.94}{$\pm$}{0.13} & \normal{8.72}{$\pm$}{0.18} & - & \secondd{7.48}{$\pm$}{0.30} & \normal{9.64}{$\pm$}{0.10} & \best{6.89}{$\pm$}{0.08} \\
& & ERGAS($\downarrow$) 
& \secondd{7.58}{$\pm$}{2.66} & \normal{16.69}{$\pm$}{6.56} & \normal{12.19}{$\pm$}{4.99} & - & \normal{8.85}{$\pm$}{2.80} & \normal{11.44}{$\pm$}{4.40} & \best{7.30}{$\pm$}{2.88}
& \normal{6.79}{$\pm$}{0.05} & \normal{6.25}{$\pm$}{0.07} & \normal{6.78}{$\pm$}{0.17} & - & \best{5.04}{$\pm$}{0.04} & \normal{7.39}{$\pm$}{0.30} & \secondd{5.16}{$\pm$}{0.07} \\
& & SSIM($\uparrow$) 
& \secondd{0.947}{$\pm$}{0.032} & \normal{0.834}{$\pm$}{0.102} & \normal{0.901}{$\pm$}{0.061} & - & \normal{0.926}{$\pm$}{0.037} & \normal{0.905}{$\pm$}{0.057} & \best{0.953}{$\pm$}{0.029}
& \normal{0.910}{$\pm$}{0.001} & \normal{0.898}{$\pm$}{0.003} & \normal{0.880}{$\pm$}{0.002} & - & \secondd{0.932}{$\pm$}{0.001} & \normal{0.883}{$\pm$}{0.005} & \best{0.935}{$\pm$}{0.002} \\
\cmidrule(lr){2-17}
% --- 32× Out-of-Dist. 数据行 ---
& \multirow{4}{*}{$\times$32}
& PSNR($\uparrow$) 
& \secondd{27.91}{$\pm$}{2.95} & \normal{21.85}{$\pm$}{2.56} & \normal{23.96}{$\pm$}{2.77} & - & \normal{27.55}{$\pm$}{2.75} & \normal{24.48}{$\pm$}{2.69} & \best{28.06}{$\pm$}{2.64}
& \normal{30.51}{$\pm$}{0.21} & \normal{29.91}{$\pm$}{0.02} & \normal{26.51}{$\pm$}{0.17} & - & \secondd{32.63}{$\pm$}{0.46} & \normal{26.53}{$\pm$}{0.41} & \best{32.79}{$\pm$}{0.26} \\
& & SAM($\downarrow$) 
& \normal{10.03}{$\pm$}{3.94} & \normal{12.26}{$\pm$}{5.90} & \normal{10.34}{$\pm$}{4.73} & - & \secondd{9.43}{$\pm$}{3.28} & \normal{9.64}{$\pm$}{3.96} & \best{8.64}{$\pm$}{3.17}
& \normal{10.26}{$\pm$}{0.51} & \normal{10.69}{$\pm$}{0.33} & \normal{10.93}{$\pm$}{0.47} & - & \secondd{9.16}{$\pm$}{0.66} & \normal{11.89}{$\pm$}{0.43} & \best{8.49}{$\pm$}{0.19} \\
& & ERGAS($\downarrow$) 
& \best{11.12}{$\pm$}{3.90} & \normal{23.54}{$\pm$}{8.22} & \normal{18.98}{$\pm$}{4.18} & - & \normal{12.35}{$\pm$}{3.30} & \normal{16.80}{$\pm$}{5.97} & \secondd{11.40}{$\pm$}{6.69}
& \normal{8.62}{$\pm$}{0.57} & \normal{7.39}{$\pm$}{0.33} & \normal{8.44}{$\pm$}{0.50} & - & \best{6.00}{$\pm$}{0.32} & \normal{9.26}{$\pm$}{0.85} & \secondd{6.24}{$\pm$}{0.30} \\
& & SSIM($\uparrow$) 
& \secondd{0.912}{$\pm$}{0.036} & \normal{0.774}{$\pm$}{0.099} & \normal{0.843}{$\pm$}{0.066} & - & \normal{0.896}{$\pm$}{0.034} & \normal{0.860}{$\pm$}{0.061} & \best{0.918}{$\pm$}{0.032}
& \normal{0.893}{$\pm$}{0.002} & \normal{0.887}{$\pm$}{0.000} & \normal{0.853}{$\pm$}{0.003} & - & \secondd{0.919}{$\pm$}{0.004} & \normal{0.856}{$\pm$}{0.001} & \best{0.922}{$\pm$}{0.001} \\
\midrule
\bottomrule
\end{tabular}%
}
\vspace{-1em}
\end{table*}

%% file: sec/4_exp.tex
\section{Experiments}
\label{sec:exp}
\subsection{Datasets}
To evaluate our model's performance and generalization, a diverse suite of seven public hyperspectral datasets are conducted, covering \textit{indoor, outdoor, and remote sensing} scenarios. Information of dataset are summarized in Tab.~\ref{tab:dataset_details}.

Our setup deliberately embraces heterogeneity, with spectral bands ranging from 31 (CAVE) to 191 (Washington DC), and substantially differing spatial resolutions and spectral ranges from sensors like ROSIS and Hyperion. This diversity provides the perfect testbed for validating the spectral agnosticism of our SSA framework. We followed standard train/test splits for each dataset, using appropriate patch sizes as detailed in the table.
% \subsubsection{Evaluation Metrics}
\subsection{Implementation Details} 
Our SSA model and all experiments were implemented using the PyTorch framework~\cite{NEURIPS2019_9015}. 
All training and inference were conducted on a single NVIDIA RTX 4090 GPU.
For the training process, we use the AdamW optimizer~\cite{loshchilov2017decoupled} and trained the model for 2000 epochs with a batch size of 4. The learning rate was managed by a cosine annealing learning rate scheduler, starting from 2e-4 and annealing to 1e-5. To enable the spectral agnosticism of MK, the crucial hyperparameter for the maximum channel, $C_{\max}$, was set to 194. Both encoders are based on the EDSR architecture~\cite{lim2017enhanced}.

\begin{figure*}
    \centering
    \includegraphics[width=0.90\textwidth]{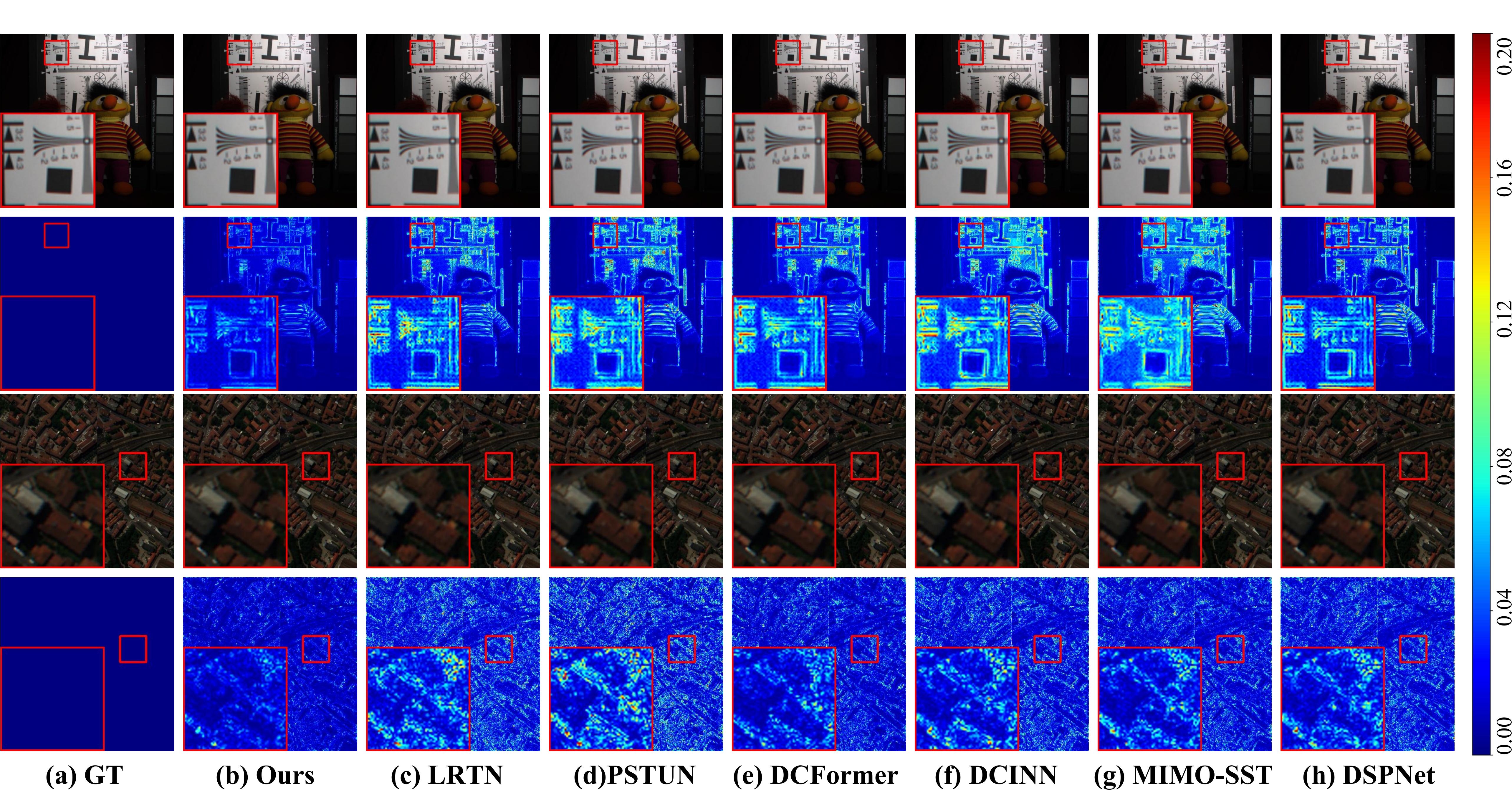}
    \caption{The upper and lower parts respectively showcase the fused images and error maps from the CAVE and PaviaC datasets. Red rectangles depict some close-up shots.}
    \label{fig:hsi}
\end{figure*}
\subsection{Main Results}
To highlight the fundamental difference between our proposed model and existing methods, for each dataset (at a $4 \times$ ratio), we trained a dedicated model independently for all SOTA methods to be compared, and evaluated them on the test sets with ratios of $4 \times$, $8 \times$, $16 \times$, and $32 \times$. For our method, we trained only a single, unified model, with the training set composed of a mixture of training samples from the seven datasets. The evaluation on the test set was conducted in the same manner as before. Additionally, we adopted four widely used quality metrics, namely Peak Signal-to-Noise Ratio (PSNR), Spectral Angle Mapper (SAM), Erreur Relative Globale Adimensionnelle de Synthèse (ERGAS), and the Structural Similarity (SSIM). The calculation formulas for the above indicators can be found in the App.\S~\ref{QualityIndexes}.
\subsubsection{Quantitative Comparison}
Tab.~\ref{tab:cave_pavia_quantitative_comparison} presents the quantitative comparison of all methods on the CAVE and PaviaC datasets, under both in-distribution (the $\times4$ training scale) and out-of-distribution ($\times8$, $\times16$, and $\times32$ scales) settings. The data in the table indicates that our method achieves SOTA performance across all in-distribution metrics. In the out-of-distribution evaluation, our model secures the best or second-best results on nearly all out-of-distribution metrics. Although methods like DCFormer~\cite{wu2025fully}, DSPNet~\cite{sun2023dual}, and PSTUN~\cite{wang2025perceptive} also achieve strong second-best results in some specific cases, they suffer from the inability to handle all scales and exhibit inconsistent performance. The quantitative comparison results for the other five datasets can be found in App.\S~\ref{sec:add_res}.
To provide a comprehensive view of the performance across all datasets, we plot the PSNR metrics of all methods in a line chart (see Fig.~\ref{fig:psnr}). It can be observed that our model does not compromise on reconstruction quality; its performance can even surpass that of specialized models that were individually trained for each specific dataset. Our method provides the best in-distribution accuracy and demonstrates robust generalization to unseen scales.
 \begin{table}[!t]
    \centering
    \caption{Ablation study of MK on the PaviaC dataset (See \S\ref{ablation:necess-nk}).}
    % Our jointly trained model with MK achieves performance on par with a hyper-specialized variant trained individually for the dataset.}
    \label{tab:ablation_mk}
    \resizebox{\columnwidth}{!}{%
    \begin{tabular}{l|c|cccc}
        \toprule
        \multicolumn{2}{c|}{\textbf{Model}} & \textbf{PSNR}($\uparrow$) & \textbf{SAM}($\downarrow$) & \textbf{ERGAS}($\downarrow$) & \textbf{SSIM}($\uparrow$) \\
        \midrule
        \multirow{2}{*}{$\times 4$} &w/o MK & 45.85 $\pm$ 1.80 & \textbf{2.01 $\pm$ 0.29} & \textbf{1.06 $\pm$ 0.11} & 0.995 $\pm$ 0.001 \\
                                         & \textbf{Ours} & \textbf{45.92 $\pm$ 1.75} & 2.02 $\pm$ 0.30 & 1.07 $\pm$ 0.11 & \textbf{0.996 $\pm$ 0.000} \\
        \hline
        \multirow{2}{*}{$\times 8$} &w/o MK & \textbf{38.62 $\pm$ 0.65} & 4.75 $\pm$ 0.07 & 3.49 $\pm$ 0.06 & 0.959 $\pm$ 0.004 \\
                                         & \textbf{Ours} & 38.59 $\pm$ 0.67 & \textbf{4.74 $\pm$ 0.06} & \textbf{3.47 $\pm$ 0.05} & \textbf{0.960 $\pm$ 0.004} \\
        \hline
        \multirow{2}{*}{$\times 16$} &w/o MK & 34.95 $\pm$ 0.03 & 6.92 $\pm$ 0.08 & 5.18 $\pm$ 0.07 & \textbf{0.936 $\pm$ 0.002} \\
                                         & \textbf{Ours} & \textbf{35.00 $\pm$ 0.02} & \textbf{6.89 $\pm$ 0.08} & \textbf{5.16 $\pm$ 0.07} & 0.935 $\pm$ 0.002 \\
        \hline
        \multirow{2}{*}{$\times 32$} &w/o MK & \textbf{32.81 $\pm$ 0.25} & 8.51 $\pm$ 0.20 & 6.25 $\pm$ 0.29 & 0.921 $\pm$ 0.001 \\
                                         & \textbf{Ours} & 32.79 $\pm$ 0.26 & \textbf{8.49 $\pm$ 0.19} & \textbf{6.24 $\pm$ 0.30} & \textbf{0.922 $\pm$ 0.001} \\
        \bottomrule
    \end{tabular}%
    }
\end{table}
\subsubsection{Qualitative Comparison}
To illustrate the advantages of our method, we provide a visual comparison in Fig.~\ref{fig:hsi} with SOTA approaches, including close-ups and error maps to highlight specific details. Visually, the results generated by our SSA model are markedly superior and most faithful to the GT. 
Visually, our SSA model yields results most faithful to the ground truth. The close-up views reveal that our method effectively recovers fine details, such as the texture of the stuffed toy and crisp lines on the chart, while competitors often suffer from blurriness and artifacts. Similarly, it reconstructs clearer architectural structures in the remote sensing scene. This superiority is corroborated by the error maps: while ours remain predominantly dark blue (indicating minimal residuals), SOTA methods display significant high-energy areas (yellow/red) along object boundaries, demonstrating the robust accuracy of our unified approach.
% To illustrate the advantages of our method, we provide a visual comparison in Fig.~\ref{fig:hsi} with six deep learning methods, including close-ups and error maps to highlight specific details. In the close-up images of ``\textit{Chart and Stuffed Toy}'' and ``\textit{ara2}'', our fusion results closely match the ground truth, achieving the highest quality. When comparing the error maps, where colors closer to blue indicate higher accuracy and colors closer to red indicate greater error, it is clear that SSA excels in detail recovery compared to other methods, whether on real outdoor remote sensing datasets or indoor hyperspectral datasets. 
\subsection{Ablation Studies}
\label{sec:ablation}
% --- Transposed Table Starts Here ---

\begin{table}[!ht]
\centering
\caption{
\textbf{Generalization to Unseen Bands and Fractional Scales.}
Top: generalization on unseen datasets at $\times4$ scale, including Houston and Loukia.
Bottom: generalization on unseen scales on CAVE (trained at $\times4$ ratio).
Results of DCFormer/PSTUN/LRTN are obtained by bicubic resizing based on $\times 4$ ratio.
}
\label{tab:generalization_fractional_unseen}
\resizebox{0.95\linewidth}{!}{%
\begin{tabular}{lcccc}
\toprule
\textbf{Model} & \textbf{PSNR($\uparrow$)} & \textbf{SAM($\downarrow$)} & \textbf{ERGAS($\downarrow$)} & \textbf{SSIM($\uparrow$)} \\
\midrule

\multicolumn{5}{c}{\textit{Unseen Dataset — Houston (48 bands, $\times4$ scale)}} \\
\midrule
DCFormer & 38.55 & 3.21 & 4.09 & 0.972 \\
PSTUN & 38.52 & 3.15 & 3.98 & 0.975 \\
LRTN & 38.65 & 3.12 & 3.91 & \textbf{0.980} \\
\textbf{Ours} & \textbf{38.71} & \textbf{3.09} & \textbf{3.85} & 0.978 \\
\midrule

\multicolumn{5}{c}{\textit{Unseen Dataset — Loukia (176 bands, $\times4$ scale)}} \\
\midrule
DCFormer & 36.25 & \textbf{3.88} & 3.65 & 0.958 \\
PSTUN & 35.80 & 4.12 & 3.82 & 0.952 \\
LRTN & 35.92 & 4.05 & 3.78 & 0.955 \\
\textbf{Ours} & \textbf{36.42} & 3.95 & \textbf{3.55} & \textbf{0.962} \\
\midrule

\multicolumn{5}{c}{\textit{Fractional Scales — CAVE (trained only at $\times4$) $\times3.2$ scale}} \\
\midrule
Bicubic & 38.50 & 5.12 & 4.85 & 0.945 \\
DCFormer & 44.65 & \textbf{2.45} & 1.95 & 0.990 \\
PSTUN & 43.90 & 2.62 & 2.25 & 0.985 \\
LRTN & 44.35 & 2.58 & 2.08 & 0.988 \\
\textbf{Ours (direct)} & \textbf{45.02} & 2.48 & \textbf{1.81} & \textbf{0.993} \\
\midrule

\multicolumn{5}{c}{\textit{Fractional Scales — CAVE (trained only at $\times4$) $\times5.7$ scale}} \\
\midrule
Bicubic & 32.20 & 6.85 & 6.10 & 0.895 \\
DCFormer & 42.95 & 3.02 & 2.45 & 0.984 \\
PSTUN & 41.25 & 3.25 & 2.72 & 0.980 \\
LRTN & 42.80 & \textbf{2.88} & 2.55 & 0.982 \\
\textbf{Ours (direct)} & \textbf{43.49} & 2.91 & \textbf{2.17} & \textbf{0.991} \\
\bottomrule
\end{tabular}}
\end{table}

\subsubsection{The Necessity of Matryoshka Kernels}
\label{ablation:necess-nk}
% --- Transposed Table Ends Here ---
To validate the role of MK, we validated an w/o MK variant by replacing our MKLs with standard fixed-channel convolutions and trained it independently on each dataset.

Tab.~\ref{tab:ablation_mk} compares this individually trained variant against our single, jointly trained model. The results show that the performance between the two models is highly comparable across all scales. While the `w/o MK' variant occasionally performs marginally better, our unified model with MK often matches or exceeds its performance.
This demonstrates that our model, achieves performance that is on par with specialized models. \\
% For further analysis, we provide more ablation studies in Suppl.\,\S 1.3, which investigates the model's generalization to unseen bands and fractional scales.
\subsubsection{Unseen Bands and Fractional Scales}
\label{sec:generalization}
We further evaluate the generalization of our pretrained model under two challenging settings: unseen datasets with different spectral bands and unseen fractional scales.

\textbf{Unseen datasets (unseen bands).}
We first test on two datasets whose spectral configurations are not observed during training: Houston with 48 bands and Loukia with 176 bands, both evaluated at the $\times4$ scale.
As shown in Tab.~\ref{tab:generalization_fractional_unseen}, our method, finetuned for only 500 iterations, consistently achieves top-tier performance across metrics.
On Houston, our model obtains the best PSNR/SAM/ERGAS (38.71\,dB / 3.09 / 3.85) and remains competitive in SSIM (0.978), demonstrating robust generalization to a different band setup.
On Loukia, our method again delivers the best reconstruction quality, achieving the highest PSNR (36.42\,dB), the lowest ERGAS (3.55), and the highest SSIM (0.962), while remaining competitive in SAM.
These results indicate that our learned spectral representation transfers well across datasets with varying band numbers and distributions.

\textbf{Unseen fractional scales.}
We further evaluate scale generalization by testing on CAVE at fractional scales $\times3.2$ and $\times5.7$, while training is performed only at the integer scale $\times4$.
For methods without native continuous scaling, we follow a strong baseline protocol: generate the $\times4$ output and then resize to the target fractional scale via Bicubic interpolation ($\times4\!\to$Bicubic).
Tab.~\ref{tab:generalization_fractional_unseen} shows that our method directly predicts fractional scales and achieves the best PSNR at both scales (45.02\,dB at $\times3.2$ and 43.49\,dB at $\times5.7$), together with the lowest ERGAS (1.81 and 2.17) and the highest SSIM (0.993 and 0.991).
Overall, these results confirm that our continuous representation provides genuine scale-agnostic capability, generalizing effectively beyond the discrete scale seen during training and outperforming strong $\times4\!\to$Bicubic baselines.

\subsubsection{Analysis of Learned Kernel Structures}
\begin{figure}[!t]
    \centering
    \includegraphics[width=1\linewidth]{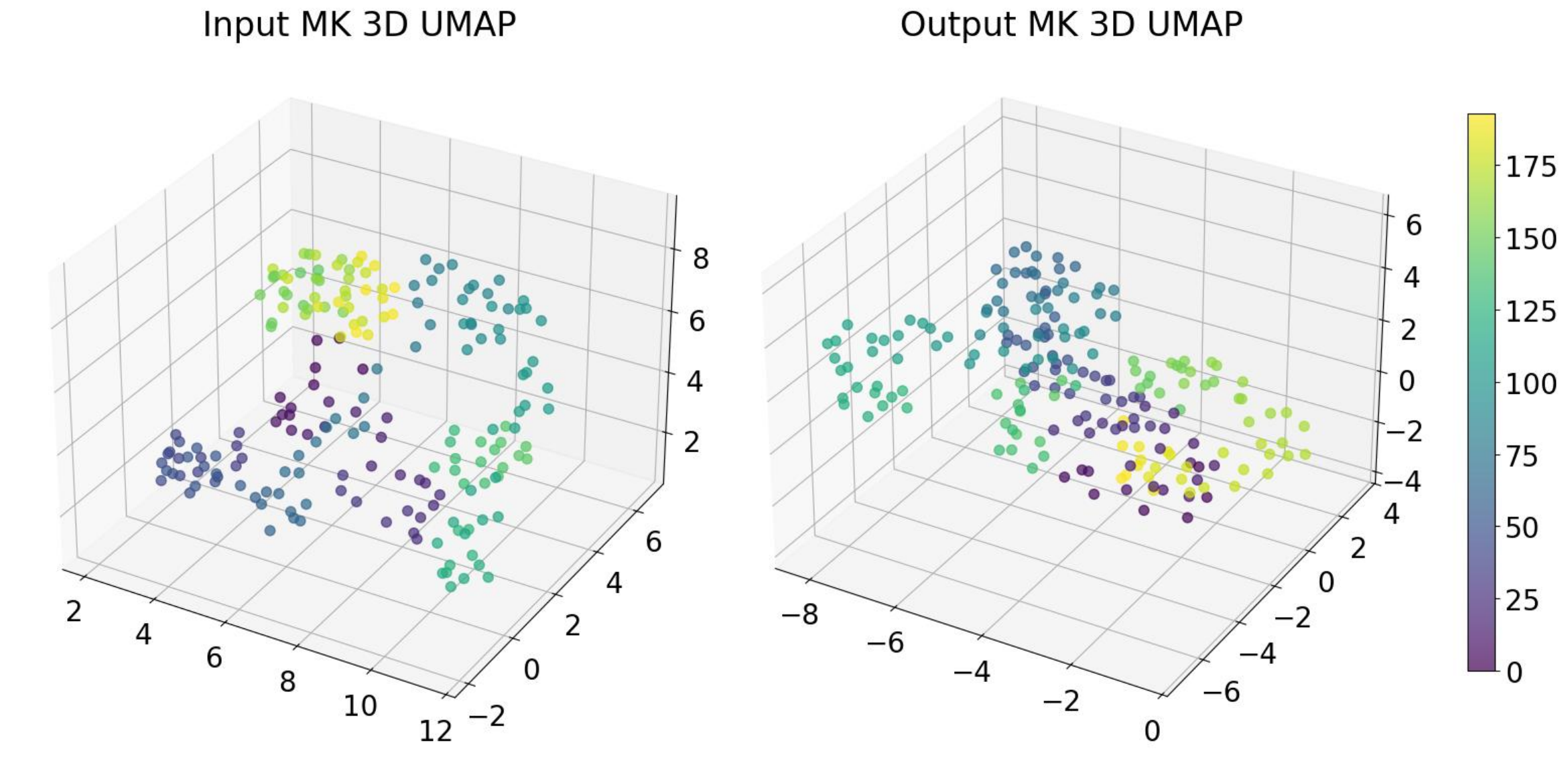}
    \caption{%Learned Structures of Input and Output Matryoshka Kernels.
     UMAP~\cite{mcinnes2018umap} visualization of the learned MK weights, which form a smooth manifold that reveals the model's encoding of spectral continuity, key to its generalization.}
    \label{fig:umap}
\end{figure}
To investigate the internal working mechanisms of our MKs, we visualized the weight structures of both the input and output kernels using UMAP~\cite{mcinnes2018umap}, as shown in Fig.~\ref{fig:umap}. Each point in the visualizations corresponds to a spectral channel, colored by its index. 
As can be seen, the projected points form smooth and continuous manifolds rather than a randomly scattered cloud. This indicates that the learning process of the kernels has formed a meaningful, layered manifold structure, where the network actively captures the high correlation between adjacent spectral bands. 
This learned internal structure is key to our model's ability to generalize effectively across different sensors.

\subsection{Efficiency Analysis}
We conducted efficiency analyses on all the comparison models and ablation experiments on our model scaling, which are detailed in App.\S~\ref{effana} and App.\S~\ref{modelscaling}.

%% file: sec/5_conclusion.tex
\section{Conclusion}
\label{sec:conclusion}
In this work, we addressed the prevalent rigidity in MS/HS fusion models by proposing SSA, a universal framework containing Matryoshka Kernels (MK) and an Implicit Neural Representation (INR) backbone overcomes the trilemma of architectural rigidity, data scarcity, and fixed spatial scaling. Our experiments demonstrate that a single, jointly trained SSA model can reach the SOTA performance, effectively generalizes to unseen sensors and scales in a zero-shot manner, and learns meaningful internal representations. While acknowledging the computational cost of INR as a potential limitation, we believe our model lays a robust foundation for the development of general-purpose foundation models in hyperspectral analysis. 

%% file: sec/X_suppl.tex
\setcounter{page}{1}

\counterwithin{table}{section}
\counterwithin{figure}{section}

\section{Matryoshka Kernel Layer Algorithms}
\label{mkl_algos}
%对于我们的MKL~\ref{subsec:mk}，其算法的伪代码如Algo.~ref{alg:adaptive_input_conv_slice}和~ref{alg:adaptive_output_conv_slice}所示。对于我们的SSA模型,我们在Matryoshka Kernel Input Layer和Matryoshka Kernel Output Layer都是用的大小为3x3的卷积核，stride、dilation、groups均为1，可以注意到我们的MKL输入输出空间大小不变。
The pseudocode for our proposed Matryoshka Kernel Layer (MKL), as detailed in \S\ref{subsec:mk}, is presented in Algo.~\ref{alg:adaptive_input_conv_slice} and Algo.~\ref{alg:adaptive_output_conv_slice}. 
In our SSA model, both the Matryoshka Kernel Input and Output Layers utilize a $3 \times 3$ convolution kernel. 
The stride, dilation, and groups are all set to 1. 
Consequently, the spatial dimensions of the input and output feature maps remain unchanged across the MKL operations.

\input{algo/inputmk}
\input{algo/outputmk}

\section{Additional Experiments Details}

\subsection{Information of Datasets}
In this section, we provide the information of used dataset, detailed in Tab.~\ref{tab:dataset_details}.

\begin{table*}[!ht]
\centering
\caption{Detailed meta information and train/test splits of the experimental datasets.}
\label{tab:dataset_details}
\resizebox{\textwidth}{!}{%
\begin{tabular}{@{}lccccccc@{}}
\toprule
\textbf{Property} & \textbf{CAVE} & \textbf{Harvard} & \textbf{PaviaC} & \textbf{PaviaU} & \textbf{Chikusei} & \textbf{Botswana} & \textbf{WashingtonDC} \\
\midrule
Bands & 31 & 31 & 102 & 103 & 128 & 145 & 191 \\
Pixel Resolution (m) & 0.001 & 0.001 & 1.3 & 1.3 & 2.5 & 30 & 2.5 \\
Spectral Range (nm) & 400--700 & 420--720 & 430--860 & 430--860 & 363--1018 & 400--2500 & 400--2500 \\
Sensors & Cooled CCD & Nuance FX & ROSIS & ROSIS & Headwall Hyperspec-VNIR-C & Hyperion & Hydice \\
Size & 512$\times$512 & 1040$\times$1392 & 1096$\times$715 & 610$\times$340 & 2517$\times$2335 & 1476$\times$256 & 1208$\times$307 \\

% Train/Test Patch Size & 128$\times$128/512$\times$512 &128$\times$128/1024$\times$1024 &128$\times$128/256$\times$256
% &128$\times$128/512$\times$512
% &128$\times$128/1024$\times$1024
% &128$\times$128/256$\times$256
% &128$\times$128/256$\times$256 \\

Train/Test Patch Size 
& \makecell{128$\times$128/ \\ 512$\times$512}
& \makecell{128$\times$128/ \\ 1024$\times$1024}
& \makecell{128$\times$128/ \\ 256$\times$256}
& \makecell{128$\times$128/ \\ 512$\times$512}
& \makecell{128$\times$128/ \\ 1024$\times$1024}
& \makecell{128$\times$128/ \\ 256$\times$256}
& \makecell{128$\times$128/ \\ 256$\times$256} \\

Train/Test Splits &3718/10 &5360/10 &2257/2 &434/2 &5175/4 &765/5  &876/5 \\

\bottomrule
\end{tabular}%
}
\end{table*}

\subsection{Data Simulation}
To comprehensively evaluate the effectiveness of our proposed method, we conducted experiments on seven public hyperspectral datasets, including two indoor/outdoor datasets, CAVE\footnote{\url{https://cave.cs.columbia.edu/repository/Multispectral}} and Harvard\footnote{\url{http://vision.seas.harvard.edu/hyperspec/index.html}}, and five remote sensing datasets: PaviaC\footnote{\url{https://www.ehu.eus/ccwintco/index.php/Hyperspectral_Remote_Sensing_Scenes##Pavia_Centre_scene}}, PaviaU\footnote{\url{https://www.ehu.eus/ccwintco/index.php/Hyperspectral_Remote_Sensing_Scenes##Pavia_University_scene}}, Chikusei\footnote{\url{https://www.sal.t.u-tokyo.ac.jp/hyperdata/}}, Botswana\footnote{\url{https://www.ehu.eus/ccwintco/index.php/Hyperspectral_Remote_Sensing_Scenes##Botswana}}, and WashingtonDC\footnote{\url{https://engineering.purdue.edu/~biehl/MultiSpec/hyperspectral.html}}. All data simulation procedures strictly comply with Wald’s Protocol~\cite{wald2000quality}. Detailed parameters for each dataset are summarized in Tab.~\ref{tab:dataset_details}. The corresponding test sets were constructed from non-overlapping image blocks, specifically four 512$\times$512 blocks for Chikusei and 256$\times$256 blocks for the other four remote sensing datasets. For all training sets, we generated 128$\times$128 image patches with a fixed stride. The low-resolution hyperspectral images (LR-HSIs) were synthesized by downsampling the original HSI by a factor of $s$ using an anti-aliasing filter, while the corresponding high-resolution multispectral images (HR-MSIs) were simulated using the spectral response functions of sensors such as ROSIS and Landsat-8.
% \subsection{Benchmark}

\subsection{Quality Indexes}
\label{QualityIndexes}

To quantitatively assess the performance of hyperspectral image reconstruction, we adopt four widely used quality indexes: PSNR, SAM, SSIM, and ERGAS. 
These metrics jointly evaluate the spatial fidelity, spectral accuracy, structural similarity, and global reconstruction consistency.

\subsubsection*{Peak Signal-to-Noise Ratio (PSNR)}

PSNR is a spatial metric that measures the ratio between the maximum possible power of a signal and the power of noise corrupting its representation. 
It provides an effective way to evaluate the spatial quality of reconstructed HSIs. 
For a reconstructed image $\widehat{\mathcal{X}}_{HR}$ and ground truth $\mathcal{X}_{HR}$, PSNR is defined as:
\begin{equation}
    PSNR(\widehat{\mathcal{X}}_{HR}, \mathcal{X}_{HR}) 
    = 20 \log \left( \frac{255}{RMSE(\widehat{\mathcal{X}}_{HR}, \mathcal{X}_{HR})} \right), \notag
\end{equation}
where RMSE denotes the root mean squared error:
\begin{equation}
    \label{eq:rmse}
    RMSE(\widehat{\mathcal{X}}_{HR}, \mathcal{X}_{HR}) 
    = \frac{ \lVert \widehat{\mathcal{X}}_{HR} - \mathcal{X}_{HR} \rVert_F }
           {\sqrt{b \cdot m_l}},
\end{equation}
in which $b$ is the number of spectral bands, and $m_l$ is the spatial pixel count. 
A higher PSNR value indicates a more accurate reconstruction.

\subsubsection*{Spectral Angle Mapper (SAM)}

SAM is a spectral metric that computes the angle between two spectral vectors. 
It is commonly used to assess spectral preservation, ensuring that the spectral 
shape is well retained during reconstruction. SAM is formulated as:
\begin{equation}
    SAM(\widehat{\mathcal{X}}_{HR}, \mathcal{X}_{HR})
    = \frac{1}{b} \sum_{j=1}^{b}
    \arccos \left(
    \frac{ \langle \widehat{x}_{j}, x_{j} \rangle }
         { \lVert \widehat{x}_{j} \rVert \lVert x_{j} \rVert }
    \right), \notag
\end{equation}
where $\widehat{x}_{j}$ and $x_{j}$ denote the reconstructed and GT spectral vectors of the $j$-th band.

\subsubsection*{Structural Similarity Index Measure (SSIM)}

SSIM evaluates the structural similarity between the reconstructed image and the ground truth. 
It jointly considers luminance, contrast, and structural information, and is widely adopted 
for perceptual image quality assessment. For each band, SSIM is defined as:
\begin{equation}
    SSIM(\widehat{x}, x)
    = \frac{ (2\mu_{\widehat{x}}\mu_x + C_1)(2\sigma_{\widehat{x}x} + C_2) }
           { (\mu_{\widehat{x}}^2 + \mu_x^2 + C_1)(\sigma_{\widehat{x}}^2 + \sigma_x^2 + C_2) }, \notag
\end{equation}
where $\mu$ and $\sigma$ denote the mean and variance, 
$\sigma_{\widehat{x}x}$ is the covariance, and $C_1$, $C_2$ are constants preventing instability.
A higher SSIM value (maximum 1) indicates better structural similarity.

\subsubsection*{Erreur Relative Globale Adimensionnelle de Synthèse (ERGAS)}

ERGAS is a global index that measures the relative reconstruction error across all spectral bands. 
Lower ERGAS values indicate better overall reconstruction performance. ERGAS is defined as:
\begin{equation}
    ERGAS(\widehat{\mathcal{X}}_{HR}, \mathcal{X}_{HR})
    = 100 \cdot \frac{1}{s}
    \sqrt{ \frac{1}{b} \sum_{j=1}^{b}
    \left( \frac{ RMSE_{j} }{ \mu_{j} } \right)^2 }, \notag
\end{equation}
where $s$ is the scale factor, $RMSE_j$ is the per-band RMSE (eq.~\ref{eq:rmse}) between 
$\widehat{\mathcal{X}}_{HR}$ and $\mathcal{X}_{HR}$, and $\mu_j$ is the mean of the GT band.

\subsection{Efficiency Analysis}
\label{effana}
Tab.~\ref{tab:efficiency} summarizes the model complexity and practical inference efficiency of representative fusion methods.
For data-driven deep learning–based hyperspectral fusion, increasing model capacity is often natural, and sometimes necessary, to effectively leverage large-scale heterogeneous training data and to improve generalization.
Accordingly, we intentionally scale up the network to enhance its fitting ability.
Even so, our model still contains fewer parameters than PSTUN~\cite{wang2025perceptive} while achieving higher reconstruction accuracy.

From the complexity perspective, Tab.~\ref{tab:efficiency} together with Tab.~\ref{tab:cave_pavia_quantitative_comparison} suggests a consistent trend:
methods with larger computational budgets generally deliver stronger reconstruction quality both in-distribution and out-of-distribution,
whereas very lightweight models (e.g., DCFormer~\cite{wu2025fully}) tend to suffer from noticeable quality degradation due to limited capacity.

Importantly, the increased capacity does not translate into prohibitive runtime.
Although our method involves higher theoretical FLOPs, it remains efficient in practice on modern GPUs:
at $\times4$ and $512\times512$ input, our model runs in 83.4\,ms per image (3.1\,MPix/s) with a peak memory footprint of 2.46\,GB.
This runtime is comparable to (or faster than) several strong baselines with similar or smaller capacity (e.g., PSTUN and LRTN),
while being orders of magnitude faster than query-based baselines such as DCINN~\cite{wang2024general}.
Overall, these results indicate that scaling the model is an effective way to improve accuracy and robustness, without sacrificing practical inference efficiency.

\begin{table}[t]
\centering
\caption{Model complexity and inference efficiency on RTX 4090.
Time/VRAM/MPix/s are measured at $\times 4$ with batch size 1 and input size $512\times512$ ($C=191$).
We report the mean runtime over 50 iterations after 20 warm-up runs using each method's validation forward entrypoint.
AMP (FP16) is enabled when supported; otherwise FP32 is used.
Peak VRAM is the maximum \texttt{cuda.max\_memory\_allocated} during inference.
Throughput is computed as $(HW/10^6)/(\text{time in seconds})$.}

\setlength{\tabcolsep}{4pt}
\renewcommand{\arraystretch}{1.15}
\small
\begin{tabular}{l
                S[table-format=3.1]
                S[table-format=4.1]
                S[table-format=3.1]
                S[table-format=2.2]
                S[table-format=3.1]}
\toprule
Method & {Params (M)$\downarrow$} & {FLOPs (G)$\downarrow$} &
{Time (ms)$\downarrow$} & {Peak VRAM (GB)$\downarrow$} & {Throughput MPix/s$\uparrow$} \\
\midrule
DSPNet~\cite{sun2023dual}   & 6.056  & 27.279 & 38.7  & 1.52 & 6.8 \\
MIMO-SST~\cite{mimo-sst}    & 4.983  & 6.164  & 14.4  & 1.22 & 18.2 \\
DCINN~\cite{wang2024general}& 2.417  & 51.531 & 860.9 & 2.25 & 0.3 \\
DCFormer~\cite{wu2025fully} & 0.082  & 0.579  & 40.2  & 1.00 & 6.5 \\
PSTUN~\cite{wang2025perceptive}& 29.948 & 102.400 & 102.4 & 1.58 & 2.6 \\
LRTN~\cite{liu2025low}      & 3.691  & 8.396  & 153.7 & 3.32 & 1.7 \\
Ours     & 18.654 & 206.848 & 83.4 & 2.46 & 3.1\\
\bottomrule
\end{tabular}
\label{tab:efficiency}
\end{table}

\begin{table}[!t]
\centering
\setlength{\tabcolsep}{1pt}
\caption{Ablation on model scaling. 
$E_{spe}$ and $E_{spa}$ denote spatial and spectral encoder depths.}
\label{tab:ablation_params}
\resizebox{0.6\linewidth}{!}{%
\begin{tabular}{ccccccc}
\toprule
$(E_{spe},E_{spa})$ & Params (M) & FLOPs & PSNR($\uparrow$)& SAM($\downarrow$) & ERGAS($\downarrow$) &SSIM($\uparrow$)\\
\midrule
$(1,1)$ & 2.735  & 58.873G & $42.45_{\pm3.24}$ & $5.32_{\pm1.20}$ & $3.92_{\pm2.05}$ & $0.985_{\pm0.004}$ \\
$(2,2)$ & 9.213  & 0.120T  & $44.15_{\pm3.12}$ & $3.94_{\pm1.05}$ & $2.88_{\pm1.50}$ & $0.988_{\pm0.003}$ \\
$(4,4)$ & 13.934 & 0.161T  & $44.72_{\pm3.04}$ & $3.05_{\pm0.95}$ & $2.11_{\pm1.28}$ & $0.991_{\pm0.002}$ \\
$(6,6)$ & 18.654 & 0.202T  & $45.96_{\pm4.69}$ & $2.55_{\pm1.00}$ & $1.72_{\pm1.13}$ & $0.994_{\pm0.005}$ \\
\bottomrule
\end{tabular}}
\end{table}

\subsection{Ablation on Model Scaling}
\label{modelscaling}
To systematically study how model capacity influences the final reconstruction quality, we conduct an ablation analysis by progressively scaling the depth of the spatial/spectral encoders. These structural changes adjust the total parameter count from 2.735M to 18.65M and FLOPs from 58.873G to 0.202T. 

As explicitly shown in Tab.~\ref{tab:ablation_params}, deploying deeper encoders consistently improves reconstruction accuracy across all metrics; for instance, the PSNR increases by over 3.5 dB from the smallest to the largest configuration. This clear trend indicates that larger models provide significantly stronger spatial–spectral representation capabilities, allowing them to better capture fine-grained features and fit the complex, high-dimensional distributions of hyperspectral data. Although this expansion naturally increases model size and computational cost, such scaling is expected and, as our results suggest, often necessary for data-driven fusion frameworks to achieve state-of-the-art performance when handling large-scale, heterogeneous hyperspectral signals.
\subsection{More Main Results}
\label{sec:add_res}
To comprehensively evaluate the universality and robustness of our proposed SSA framework, we extend our experiments to five additional benchmarks: Harvard, PaviaU, Chikusei, Botswana, and WashingtonDC. These datasets encompass a wide variety of sensor characteristics, with spectral bands ranging from 31 to 191. As presented in Tabs.~\ref{tab:harvard_multi_scales_singlecell} through \ref{tab:washingtondc_multi_scales_right_align}, our unified model consistently achieves state-of-the-art performance across these diverse scenarios. Remarkably, even though our universal model is trained jointly on heterogeneous datasets, it outperforms specialized approaches like LRTN~\cite{liu2025low} and PSTUN~\cite{wang2025perceptive} that are trained individually for each specific dataset. For instance, in the in-distribution setting ($\times 4$ scale) on the WashingtonDC dataset (Tab.~\ref{tab:washingtondc_multi_scales_right_align}), our method surpasses the second-best approach by approximately 1.0 dB in PSNR. This result strongly validates that our Matryoshka Kernel design effectively mitigates the interference typically associated with multi-dataset training, allowing a single network to adapt to varying spectral distributions without architectural modification.

\begin{figure*}[!b]
    \centering
    \includegraphics[width=0.95\textwidth]{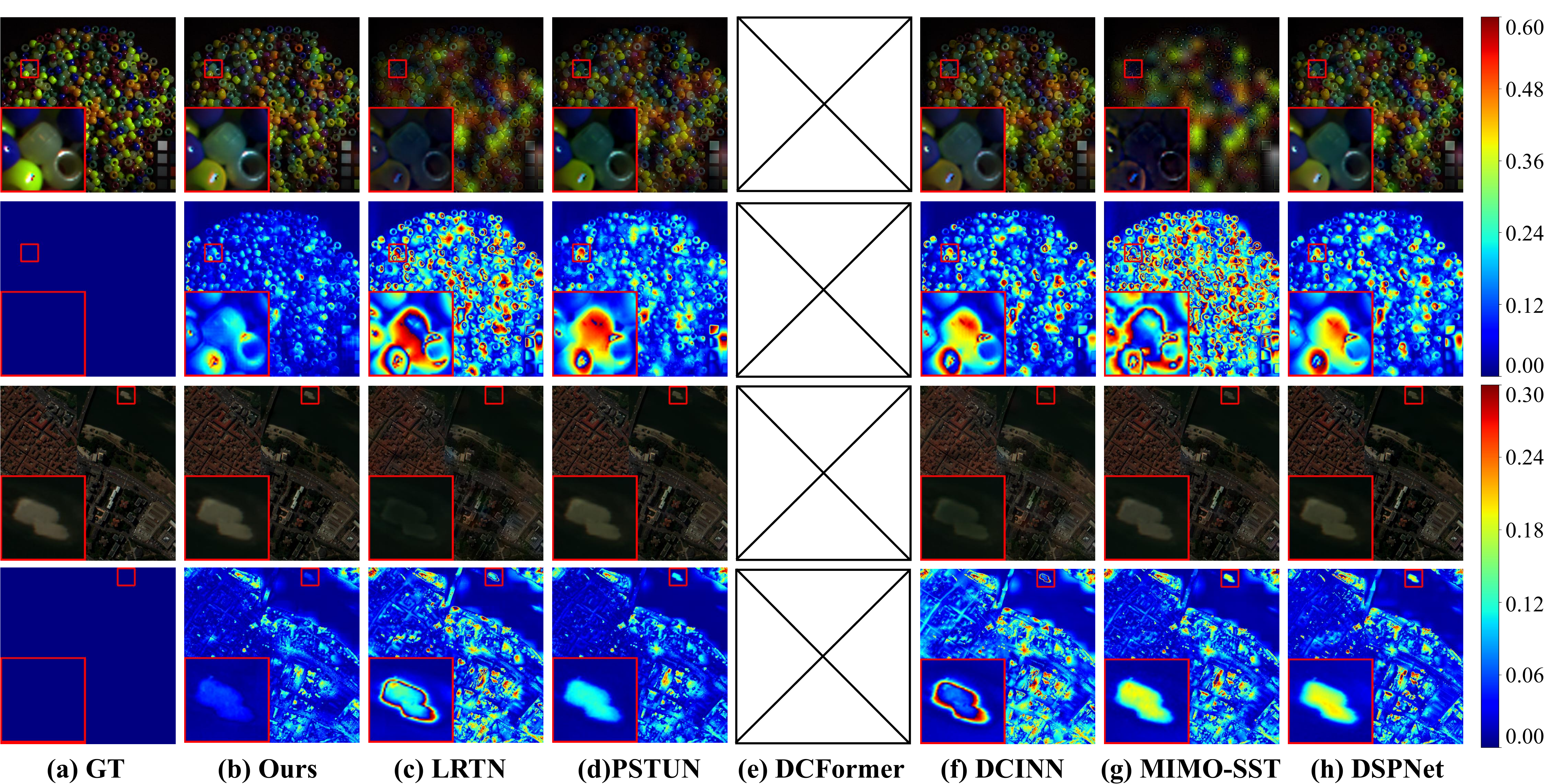}
    \caption{Qualitative comparison on the \textbf{CAVE} (top rows) and \textbf{PaviaC} (bottom rows) datasets at the unseen \textbf{$\times 32$ scale}. The red boxes highlight close-up details, and the even rows display the corresponding error maps. Our SSA model exhibits superior detail recovery and minimal reconstruction error compared to SOTA methods, which suffer from significant blurriness at this extreme scale.}
    \label{fig:cave_paviac_vis}
\end{figure*}

\begin{figure*}
    \centering
    \includegraphics[width=0.95\textwidth]{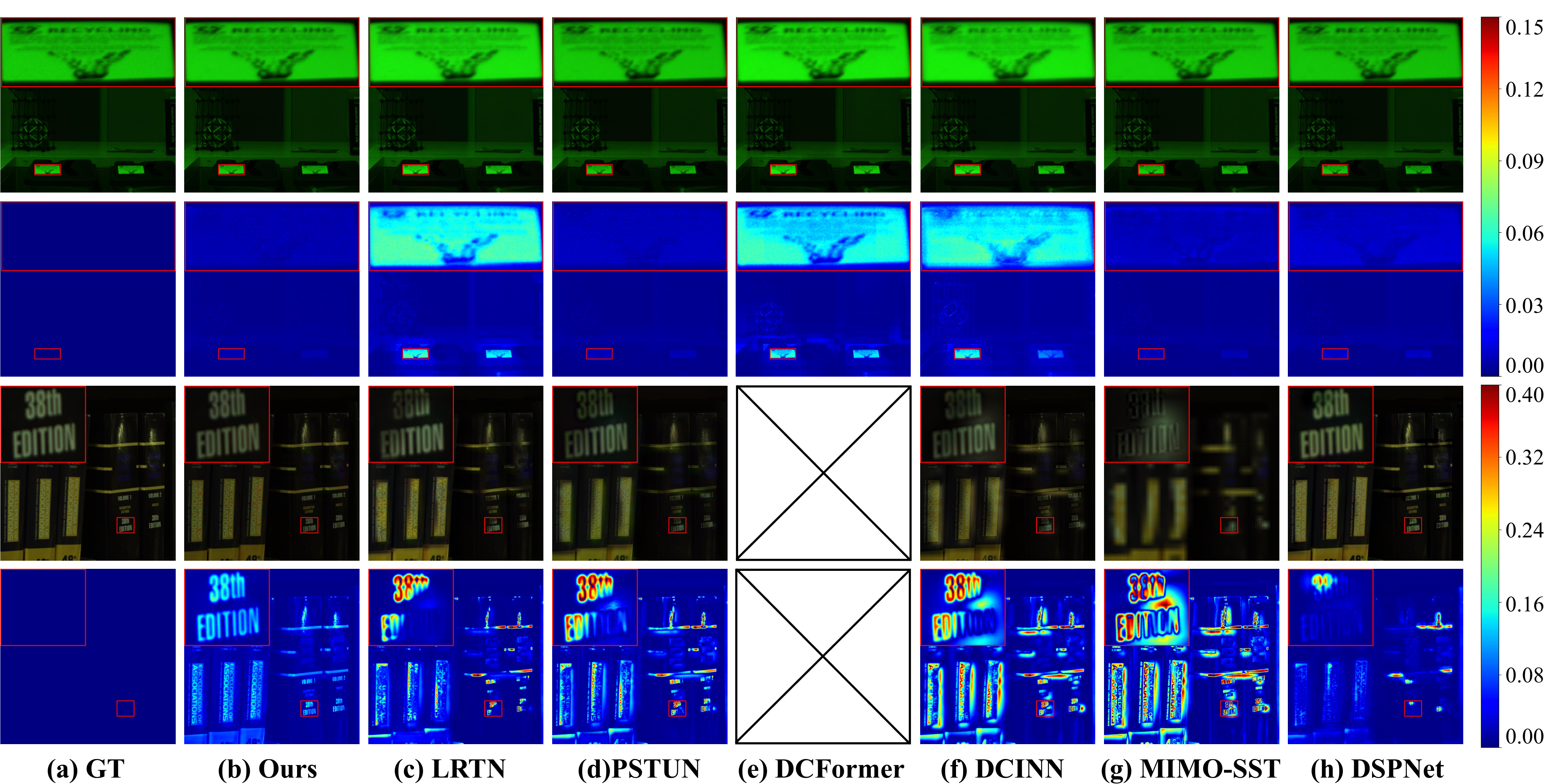}
    \caption{Visual comparison on the \textbf{Harvard} dataset. The top two rows correspond to the in-distribution $\times 4$ scale, and the bottom two rows correspond to the unseen $\times 32$ scale. Our method preserves sharp textual details (see zoom-in) even at large magnification factors.}
    \label{fig:harvard_vis}
\end{figure*}

\begin{figure*}
    \centering
    \includegraphics[width=0.95\textwidth]{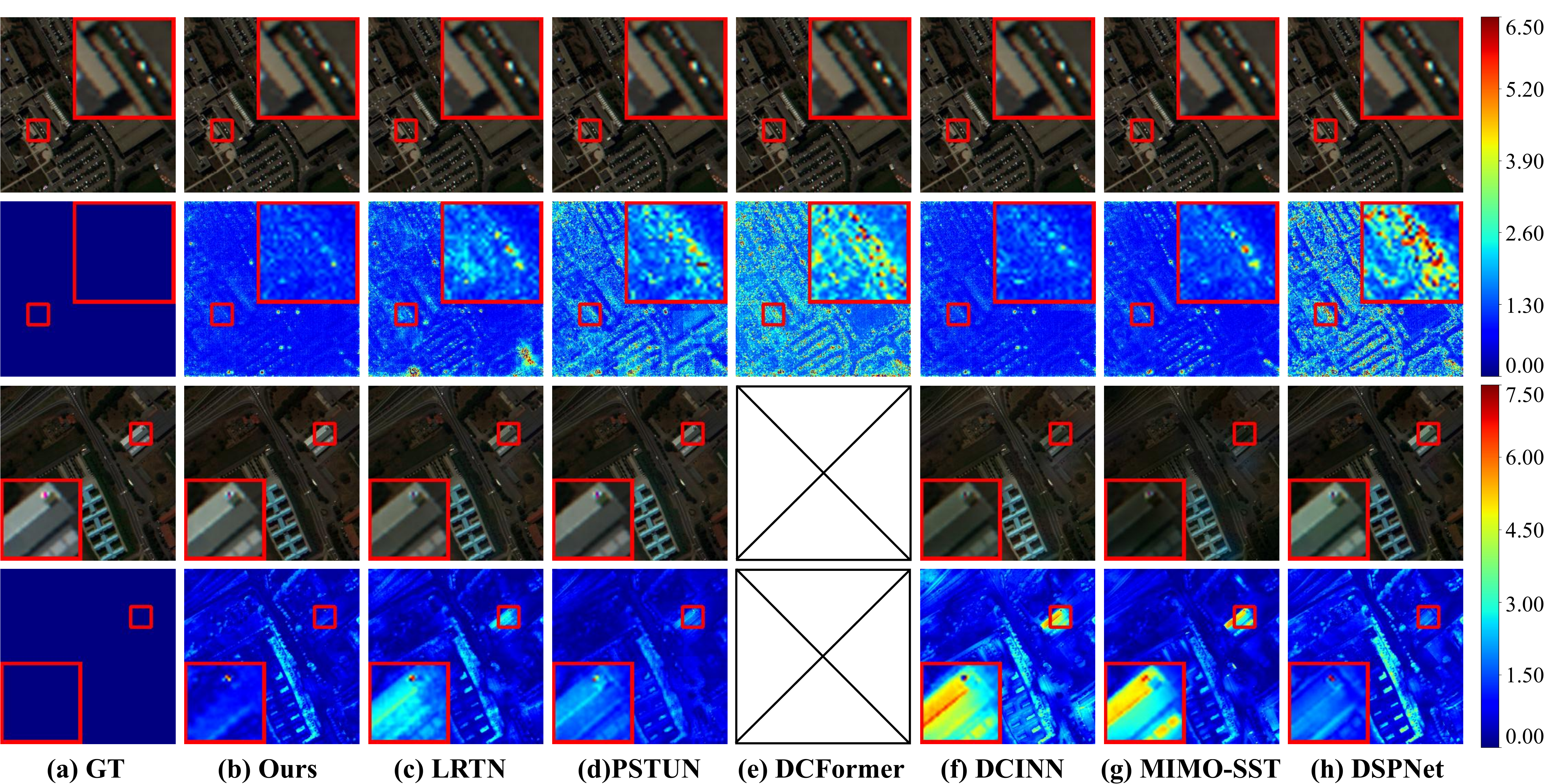}
    \caption{Visual comparison on the \textbf{PaviaU} dataset across varying scales. The top two rows show the results at the training scale ($\times 4$), while the bottom two rows display the out-of-distribution scale ($\times 32$). For each scale, we present the fused image and its corresponding error map. Note that DCFormer~\cite{wu2025fully} fails to support the arbitrary $\times 32$ scale (indicated by crossed boxes).}
    \label{fig:paviau_vis}
\end{figure*}

\begin{figure*}
    \centering
    \includegraphics[width=0.95\textwidth]{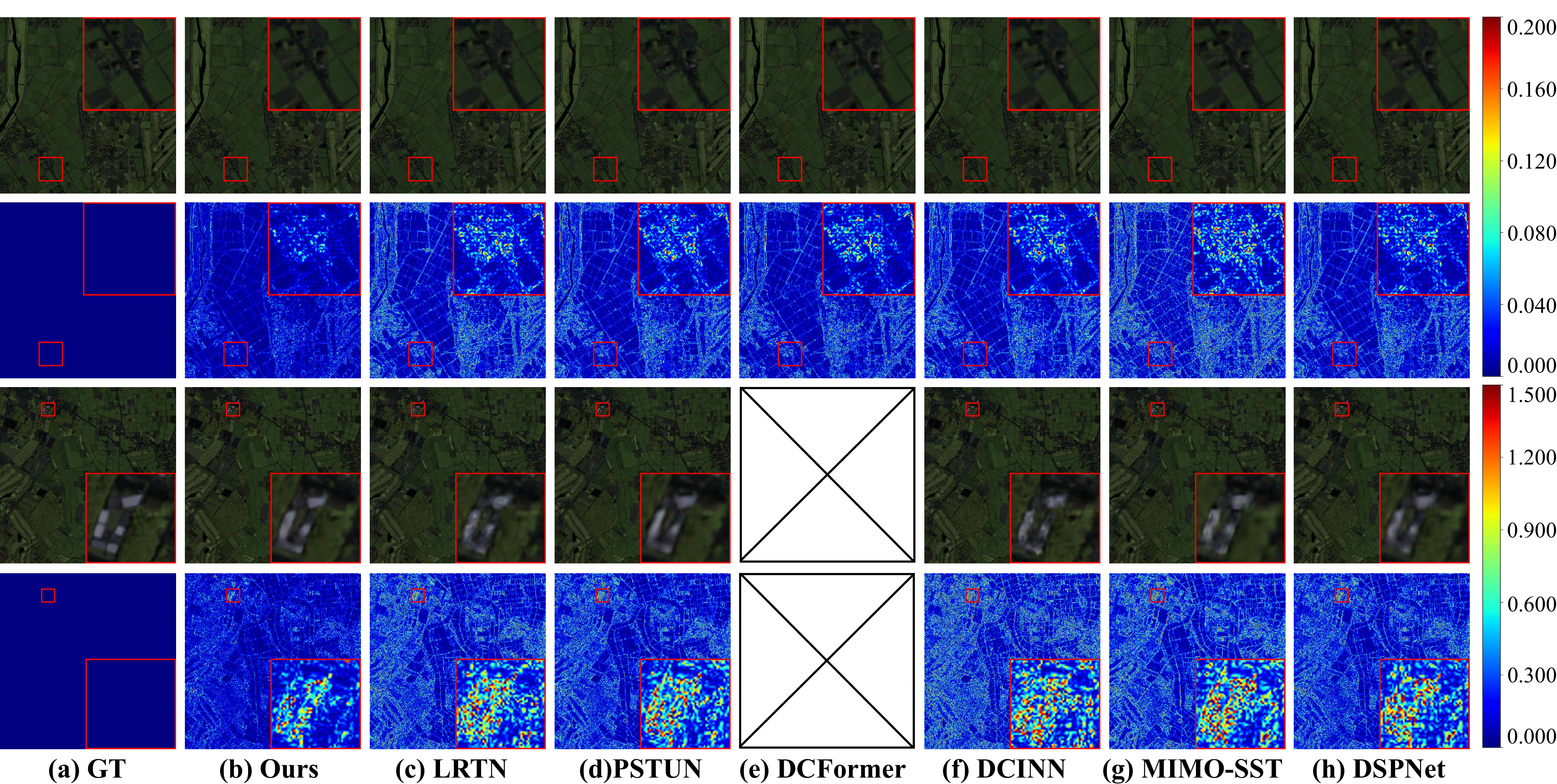}
    \caption{Visual comparison on the \textbf{Chikusei} dataset. Top two rows: $\times 4$ scale; Bottom two rows: $\times 32$ scale. The error maps (blue indicates low error) demonstrate that our model generalizes well to large-scale remote sensing scenes with varying spatial resolutions.}
    \label{fig:chikusei_vis}
\end{figure*}

\begin{figure*}
    \centering
    \includegraphics[width=0.95\textwidth]{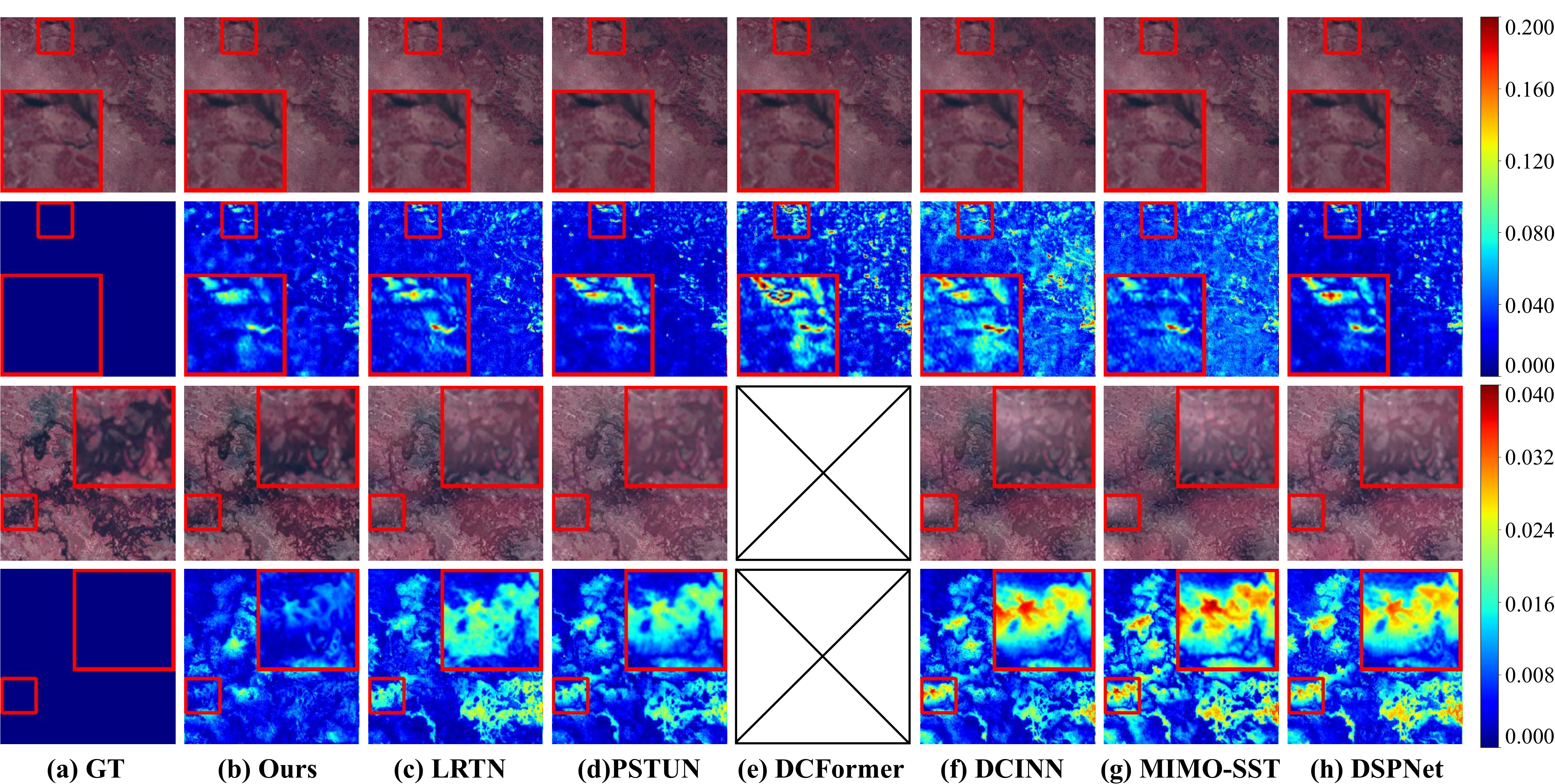}
    \caption{Visual comparison on the \textbf{Botswana} dataset. Top two rows: $\times 4$ scale; Bottom two rows: $\times 32$ scale. Our approach maintains spectral and spatial consistency better than comparison methods, which often show high-energy residuals (red/yellow) at large upscaling factors.}
    \label{fig:botswana_vis}
\end{figure*}

\begin{figure*}
    \centering
    \includegraphics[width=0.95\textwidth]{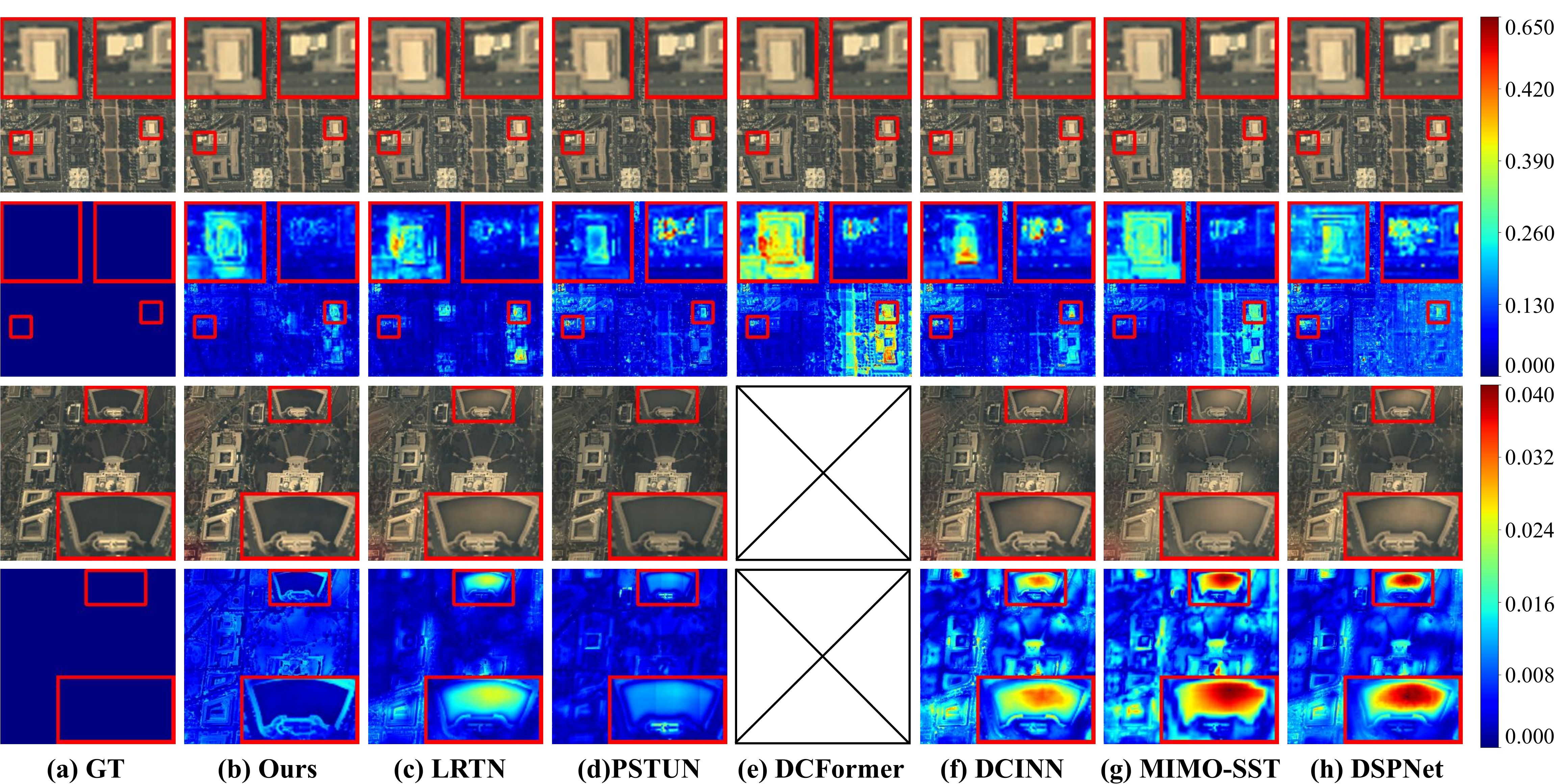}
    \caption{Visual comparison on the \textbf{WashingtonDC} dataset. Top two rows: $\times 4$ scale; Bottom two rows: $\times 32$ scale. Despite the high band count (191 bands), our unified model robustly reconstructs spatial details across different scaling factors without model retraining.}
    \label{fig:washingtondc_vis}
\end{figure*}

We further corroborate these quantitative findings with comprehensive visual comparisons provided in Figs.~\ref{fig:cave_paviac_vis} through \ref{fig:washingtondc_vis}. To explicitly demonstrate the scale generalization capability of our method, we structured the visualizations for the five additional datasets (Figs.~\ref{fig:paviau_vis}--\ref{fig:washingtondc_vis}) to display the training scale ($\times 4$, top two rows) and the extreme unseen scale ($\times 32$, bottom two rows) side-by-side. 
For the CAVE and PaviaC datasets, we specifically showcase the challenging $\times 32$ results in Fig.~\ref{fig:cave_paviac_vis}.
As observed, while most competitive methods maintain reasonable quality at the $\times 4$ scale, their performance degrades significantly at $\times 32$. Methods with fixed upsampling modules, such as DCFormer~\cite{wu2025fully}, completely fail to generate valid outputs at this unseen resolution (indicated by crossed-out boxes), and others exhibit severe blurring or grid-like artifacts. In contrast, leveraging the continuous modeling capability of our INR backbone, SSA maintains high reconstruction fidelity with predominantly dark blue error maps across all scales. Notably, in the Harvard dataset (Fig.~\ref{fig:harvard_vis}), our model successfully recovers legible text (e.g., ``30th EDITION'') even at the $\times 32$ scale, whereas other methods suffer from heavy distortions.

\begin{table*}[!h]
\centering
\caption{Quantitative comparison on the \textbf{Harvard} dataset across four scales. Values are mean\,$\pm$\, std in a single cell.}
\label{tab:harvard_multi_scales_singlecell}
\resizebox{\textwidth}{!}{%
\begin{tabular}{@{}lll*7{c}@{}}
\toprule
\midrule
\multicolumn{3}{c}{\textbf{Metric}} &
\multicolumn{7}{c}{\textbf{Harvard Dataset}} \\
\cmidrule(l){4-10}
\multicolumn{3}{c}{} & \textbf{DSPNet} & \textbf{MIMO-SST} & \textbf{DCINN} & \textbf{DCFormer} & \textbf{PSTUN} & \textbf{LRTN} & \textbf{Ours} \\
\midrule
% ========================= ×4 In-Dist. =========================
\multirow{4}{*}{\rotatebox[origin=c]{90}{In-Dist.}} &
\multirow{4}{*}{$\times$4} &
PSNR($\uparrow$)   &
\normal{43.97}{$\pm$}{5.99} &
\normal{43.65}{$\pm$}{6.85} &
\normal{42.06}{$\pm$}{4.55} &
\normal{42.87}{$\pm$}{5.30} &
\secondd{44.30}{$\pm$}{4.59} &
\normal{41.91}{$\pm$}{5.33} &
\best{44.81}{$\pm$}{5.21} \\
& & SAM($\downarrow$)  &
\best{2.70}{$\pm$}{0.83} &
\normal{2.98}{$\pm$}{0.86} &
\normal{2.96}{$\pm$}{0.85} &
\normal{3.19}{$\pm$}{0.91} &
\normal{3.59}{$\pm$}{0.98} &
\normal{4.58}{$\pm$}{1.54} &
\secondd{2.73}{$\pm$}{0.89} \\
& & ERGAS($\downarrow$) &
\best{4.30}{$\pm$}{2.29} &
\normal{5.11}{$\pm$}{2.08} &
\normal{5.46}{$\pm$}{0.90} &
\normal{5.12}{$\pm$}{2.46} &
\normal{4.54}{$\pm$}{1.80} &
\normal{5.78}{$\pm$}{1.38} &
\secondd{4.41}{$\pm$}{2.63} \\
& & SSIM($\uparrow$)   &
\secondd{0.983}{$\pm$}{0.001} &
\normal{0.983}{$\pm$}{0.007} &
\normal{0.983}{$\pm$}{0.003} &
\normal{0.982}{$\pm$}{0.009} &
\normal{0.983}{$\pm$}{0.008} &
\normal{0.975}{$\pm$}{0.008} &
\best{0.984}{$\pm$}{0.010} \\
\cmidrule(lr){1-10}
% ========================= ×8/16/32 Out-of-Dist. =========================
\multirow{12}{*}{\rotatebox[origin=c]{90}{Out-of-Dist.}} &
\multirow{4}{*}{$\times$8} &
PSNR($\uparrow$)   &
\normal{42.25}{$\pm$}{6.20} &
\normal{38.62}{$\pm$}{6.74} &
\normal{40.61}{$\pm$}{6.23} &
\secondd{42.37}{$\pm$}{5.25} &
\normal{41.79}{$\pm$}{4.04} &
\normal{41.33}{$\pm$}{5.63} &
\best{42.73}{$\pm$}{5.09} \\
& & SAM($\downarrow$)  &
\best{2.99}{$\pm$}{0.90} &
\normal{3.26}{$\pm$}{0.98} &
\normal{3.33}{$\pm$}{0.90} &
\normal{3.51}{$\pm$}{0.93} &
\normal{4.00}{$\pm$}{0.99} &
\normal{4.82}{$\pm$}{1.63} &
\secondd{3.02}{$\pm$}{0.92} \\
& & ERGAS($\downarrow$) &
\secondd{5.19}{$\pm$}{2.69} &
\normal{6.68}{$\pm$}{2.43} &
\normal{5.69}{$\pm$}{2.51} &
\normal{5.30}{$\pm$}{2.15} &
\normal{5.62}{$\pm$}{1.84} &
\normal{5.98}{$\pm$}{1.23} &
\best{4.92}{$\pm$}{2.08} \\
& & SSIM($\uparrow$)   &
\normal{0.976}{$\pm$}{0.012} &
\normal{0.944}{$\pm$}{0.047} &
\normal{0.974}{$\pm$}{0.012} &
\best{0.980}{$\pm$}{0.012} &
\normal{0.977}{$\pm$}{0.008} &
\normal{0.972}{$\pm$}{0.013} &
\secondd{0.978}{$\pm$}{0.013} \\
\cmidrule(lr){2-10}
& \multirow{4}{*}{$\times$16} &
PSNR($\uparrow$)   &
\normal{37.31}{$\pm$}{6.99} &
\normal{34.36}{$\pm$}{7.28} &
\normal{35.94}{$\pm$}{6.79} &
-- &
\normal{37.46}{$\pm$}{5.23} &
\secondd{38.10}{$\pm$}{6.48} &
\best{38.73}{$\pm$}{6.00} \\
& & SAM($\downarrow$)  &
\best{3.60}{$\pm$}{1.15} &
\normal{4.01}{$\pm$}{1.25} &
\normal{3.95}{$\pm$}{1.11} &
-- &
\normal{4.89}{$\pm$}{1.24} &
\normal{5.05}{$\pm$}{1.56} &
\secondd{3.85}{$\pm$}{1.00} \\
& & ERGAS($\downarrow$) &
\best{7.53}{$\pm$}{4.48} &
\normal{10.74}{$\pm$}{4.12} &
\normal{8.90}{$\pm$}{2.98} &
-- &
\normal{9.40}{$\pm$}{5.05} &
\normal{8.56}{$\pm$}{4.39} &
\secondd{7.68}{$\pm$}{5.21} \\
& & SSIM($\uparrow$)   &
\secondd{0.956}{$\pm$}{0.031} &
\normal{0.903}{$\pm$}{0.069} &
\normal{0.950}{$\pm$}{0.024} &
-- &
\normal{0.951}{$\pm$}{0.036} &
\normal{0.952}{$\pm$}{0.036} &
\best{0.960}{$\pm$}{0.037} \\
\cmidrule(lr){2-10}
& \multirow{4}{*}{$\times$32} &
PSNR($\uparrow$)   &
\normal{36.73}{$\pm$}{6.72} &
\normal{31.39}{$\pm$}{7.65} &
\normal{31.78}{$\pm$}{7.14} &
-- &
\normal{35.78}{$\pm$}{4.84} &
\secondd{36.97}{$\pm$}{6.37} &
\best{37.34}{$\pm$}{6.49} \\
& & SAM($\downarrow$)  &
\best{4.28}{$\pm$}{1.38} &
\normal{5.63}{$\pm$}{1.56} &
\normal{4.99}{$\pm$}{1.27} &
-- &
\normal{5.94}{$\pm$}{1.81} &
\normal{5.68}{$\pm$}{1.91} &
\secondd{4.66}{$\pm$}{1.57} \\
& & ERGAS($\downarrow$) &
\secondd{9.21}{$\pm$}{5.44} &
\normal{13.41}{$\pm$}{5.37} &
\normal{12.46}{$\pm$}{4.68} &
-- &
\normal{11.57}{$\pm$}{5.72} &
\normal{9.50}{$\pm$}{4.14} &
\best{8.70}{$\pm$}{4.79} \\
& & SSIM($\uparrow$)   &
\normal{0.942}{$\pm$}{0.044} &
\normal{0.863}{$\pm$}{0.090} &
\normal{0.902}{$\pm$}{0.053} &
-- &
\normal{0.941}{$\pm$}{0.033} &
\secondd{0.945}{$\pm$}{0.038} &
\best{0.949}{$\pm$}{0.045} \\
\midrule
\bottomrule
\end{tabular}%
}
\end{table*}
\input{Tabs/tab_harvard}



%% file: algo/inputmk.tex
\begin{algorithm}
\caption{Matryoshka Kernel Input Layer Forward Pass}
\label{alg:adaptive_input_conv_slice}
\begin{algorithmic}[1]

\REQUIRE
\INPUT $\mathbf{X} \in \mathbb{R}^{N \times C_{\text{in}} \times H \times W}$
\INPUT Nested convolution weight 
$\mathbf{W}_{\text{nested}} \in \mathbb{R}^{D \times C_{\text{max}} \times k \times k}$
\INPUT Convolution bias $\mathbf{b} \in \mathbb{R}^{D}$ (optional)
\INPUT Convolution parameters: stride $\mathbf{s}$, padding $\mathbf{p}$,
dilation $\mathbf{d}$, groups $\mathbf{g}$

\ENSURE
\OUTPUT $\mathbf{Y} \in \mathbb{R}^{N \times D \times H' \times W'}$

\FUNCTION{Forward$(\mathbf{X}, \mathbf{W}_{\text{nested}}, \mathbf{b})$}

\IF{$C_{\text{in}} = C_{\text{max}}$}
    \STATE $\mathbf{Y} \gets
    \texttt{Conv2d}(\mathbf{X}, \mathbf{W}_{\text{nested}}, \mathbf{b},
    \mathbf{s}, \mathbf{p}, \mathbf{d}, \mathbf{g})$
    \COMMENT{Use full weights}
\ELSE
    \STATE $\mathbf{W}_{\text{valid}} \gets
    \mathbf{W}_{\text{nested}}[:, :C_{\text{in}}, :, :]$
    \COMMENT{Slice weights for input channels $C_{\text{in}}$}

    \STATE $\mathbf{Y} \gets
    \texttt{Conv2d}(\mathbf{X}, \mathbf{W}_{\text{valid}}, \mathbf{b},
    \mathbf{s}, \mathbf{p}, \mathbf{d}, \mathbf{g})$
\ENDIF

\STATE \textbf{return} $\mathbf{Y}$

\ENDFUNCTION

\end{algorithmic}
\end{algorithm}

%% file: algo/outputmk.tex
\begin{algorithm}
\caption{Matryoshka Kernel Output Layer Forward Pass}
\label{alg:adaptive_output_conv_slice}
\begin{algorithmic}[1]

\REQUIRE
\INPUT $\mathbf{Y} \in \mathbb{R}^{N \times D \times H' \times W'}$
\INPUT Target output channels $C_{\text{out}}$
\INPUT Pre-defined convolution weight
$\mathbf{W}_{\text{nested}} \in \mathbb{R}^{C_{\text{max}} \times D \times k \times k}$
\INPUT Pre-defined convolution bias
$\mathbf{b}_{\text{nested}} \in \mathbb{R}^{C_{\text{max}}}$ (optional)
\INPUT Convolution parameters: stride $\mathbf{s}$, padding $\mathbf{p}$,
dilation $\mathbf{d}$, groups $\mathbf{g}$

\ENSURE
\OUTPUT $\mathbf{X} \in \mathbb{R}^{N \times C_{\text{out}} \times H \times W}$

\FUNCTION{Forward$(\mathbf{Y}, C_{\text{out}}, \mathbf{W}_{\text{nested}}, \mathbf{b}_{\text{nested}})$}

\IF{$C_{\text{out}} = C_{\text{max}}$}
    \STATE $\mathbf{X} \gets
    \texttt{Conv2d}(\mathbf{Y}, \mathbf{W}_{\text{nested}},
    \mathbf{b}_{\text{nested}}, \mathbf{s}, \mathbf{p}, \mathbf{d}, \mathbf{g})$
\ELSE
    \STATE $\mathbf{W}_{\text{valid}} \gets
    \mathbf{W}_{\text{nested}}[:C_{\text{out}}, :, :, :]$
    \COMMENT{Slice weights for output channels $C_{\text{out}}$}

    \STATE $\mathbf{b}_{\text{valid}} \gets \text{None}$

    \IF{$\mathbf{b}_{\text{nested}} \neq \text{None}$}
        \STATE $\mathbf{b}_{\text{valid}} \gets
        \mathbf{b}_{\text{nested}}[:C_{\text{out}}]$
        \COMMENT{Slice the bias}
    \ENDIF

    \STATE $\mathbf{X} \gets
    \texttt{Conv2d}(\mathbf{Y}, \mathbf{W}_{\text{valid}},
    \mathbf{b}_{\text{valid}}, \mathbf{s}, \mathbf{p}, \mathbf{d}, \mathbf{g})$
\ENDIF

\STATE \textbf{return} $\mathbf{X}$

\ENDFUNCTION

\end{algorithmic}
\end{algorithm}

%% file: Tabs/tab_harvard.tex
\begin{table*}[t]
\centering
\caption{Quantitative comparison on the \textbf{PaviaU} dataset across four scales. Values are mean\,$\pm$\,std.}
\label{tab:paviau_multi_scales_right_align}
\sisetup{}
\resizebox{\textwidth}{!}{%
% 3 列描述 + 7 个方法，每个方法 1 列（单元格内 perfalign）
\begin{tabular}{@{}lll*7{c}@{}}
\toprule
\midrule
\multicolumn{3}{c}{\textbf{Metric}} &
\multicolumn{7}{c}{\textbf{PaviaU Dataset}} \\
\cmidrule(l){4-10}
\multicolumn{3}{c}{} & \textbf{DSPNet} & \textbf{MIMO-SST} & \textbf{DCINN} & \textbf{DCFormer} & \textbf{PSTUN} & \textbf{LRTN} & \textbf{Ours} \\
\midrule
% ========================= ×4 In-Dist. =========================
\multirow{4}{*}{\rotatebox[origin=c]{90}{In-Dist.}} &
\multirow{4}{*}{$\times$4} &
PSNR($\uparrow$)   &
\normal{37.51}{$\pm$}{0.15} &
\normal{40.28}{$\pm$}{0.15} &
\secondd{40.31}{$\pm$}{0.42} &
\normal{39.05}{$\pm$}{0.25} &
\normal{39.46}{$\pm$}{0.17} &
\normal{40.09}{$\pm$}{0.63} &
\best{40.44}{$\pm$}{0.15} \\
& & SAM($\downarrow$)  &
\normal{2.27}{$\pm$}{0.12} &
\normal{3.46}{$\pm$}{0.11} &
\secondd{1.96}{$\pm$}{0.07} &
\normal{2.60}{$\pm$}{0.17} &
\normal{2.54}{$\pm$}{0.02} &
\normal{2.05}{$\pm$}{0.14} &
\best{1.95}{$\pm$}{0.07} \\
& & ERGAS($\downarrow$)&
\normal{1.99}{$\pm$}{0.07} &
\normal{1.85}{$\pm$}{0.11} &
\best{1.53}{$\pm$}{0.02} &
\normal{1.77}{$\pm$}{0.07} &
\normal{1.58}{$\pm$}{0.02}  &
\secondd{1.55}{$\pm$}{0.00} &
\normal{1.56}{$\pm$}{0.06}\\
& & SSIM($\uparrow$)   &
\normal{0.981}{$\pm$}{0.001} &
\normal{0.977}{$\pm$}{0.002} &
\best{0.988}{$\pm$}{0.001} &
\normal{0.977}{$\pm$}{0.000} &
\normal{0.981}{$\pm$}{0.000} &
\secondd{0.987}{$\pm$}{0.001} &
\best{0.988}{$\pm$}{0.001} \\
\cmidrule(lr){1-10}
% ========================= ×8/16/32 Out-of-Dist. =========================
\multirow{12}{*}{\rotatebox[origin=c]{90}{Out-of-Dist.}} &
\multirow{4}{*}{$\times$8} &
PSNR($\uparrow$)   &
\normal{35.76}{$\pm$}{0.50} &
\normal{31.43}{$\pm$}{0.55} &
\normal{28.91}{$\pm$}{0.44} &
\normal{30.56}{$\pm$}{1.74} &
\normal{36.01}{$\pm$}{0.26} &
\secondd{36.18}{$\pm$}{0.03} &
\best{36.68}{$\pm$}{0.26} \\
& & SAM($\downarrow$)  &
\normal{5.57}{$\pm$}{0.06} &
\normal{5.83}{$\pm$}{0.17} &
\normal{5.36}{$\pm$}{0.26} &
\normal{5.39}{$\pm$}{0.24} &
\best{4.79}{$\pm$}{0.27} &
\normal{5.16}{$\pm$}{0.20} &
\secondd{4.97}{$\pm$}{0.30} \\
& & ERGAS($\downarrow$)&
\normal{4.64}{$\pm$}{0.01} &
\normal{4.47}{$\pm$}{0.12} &
\normal{4.89}{$\pm$}{0.19} &
\normal{4.06}{$\pm$}{0.56} &
\secondd{3.50}{$\pm$}{0.15} &
\normal{4.01}{$\pm$}{0.14}  &
\best{3.03}{$\pm$}{0.06}\\
& & SSIM($\uparrow$)   &
\normal{0.917}{$\pm$}{0.001} &
\normal{0.895}{$\pm$}{0.001} &
\normal{0.884}{$\pm$}{0.003} &
\normal{0.927}{$\pm$}{0.004} &
\secondd{0.959}{$\pm$}{0.002} &
\normal{0.940}{$\pm$}{0.001} &
\best{0.960}{$\pm$}{0.002} \\
\cmidrule(lr){2-10}
& \multirow{4}{*}{$\times$16} &
PSNR($\uparrow$)   &
\normal{32.76}{$\pm$}{0.79} &
\normal{27.29}{$\pm$}{1.12} &
\normal{26.47}{$\pm$}{0.56} &
-- &
\secondd{33.66}{$\pm$}{0.49} &
\normal{33.26}{$\pm$}{0.00} &
\best{35.48}{$\pm$}{0.16} \\
& & SAM($\downarrow$)  &
\normal{7.63}{$\pm$}{0.31} &
\normal{9.18}{$\pm$}{0.16} &
\normal{8.37}{$\pm$}{0.34} &
-- &
\secondd{6.35}{$\pm$}{0.14}  &
\normal{6.38}{$\pm$}{0.23} &
\best{6.26}{$\pm$}{0.34}\\
& & ERGAS($\downarrow$)&
\normal{5.72}{$\pm$}{0.47} &
\normal{7.01}{$\pm$}{0.78} &
\normal{6.51}{$\pm$}{0.25} &
-- &
\normal{5.62}{$\pm$}{0.01}  &
\secondd{5.03}{$\pm$}{0.15} &
\best{4.40}{$\pm$}{0.06}\\
& & SSIM($\uparrow$)   &
\normal{0.873}{$\pm$}{0.004} &
\normal{0.844}{$\pm$}{0.005} &
\normal{0.833}{$\pm$}{0.005} &
-- &
\normal{0.840}{$\pm$}{0.003}  &
\secondd{0.879}{$\pm$}{0.003} &
\best{0.912}{$\pm$}{0.000}\\
\cmidrule(lr){2-10}
& \multirow{4}{*}{$\times$32} &
PSNR($\uparrow$)   &
\normal{30.50}{$\pm$}{1.66} &
\normal{24.39}{$\pm$}{1.25} &
\normal{24.25}{$\pm$}{0.21} &
-- &
\secondd{31.49}{$\pm$}{0.64} &
\normal{30.44}{$\pm$}{0.18} &
\best{34.35}{$\pm$}{0.14} \\
& & SAM($\downarrow$)  &
\normal{10.50}{$\pm$}{0.16} &
\normal{12.63}{$\pm$}{0.87} &
\normal{11.47}{$\pm$}{0.49} &
-- &
\normal{9.17}{$\pm$}{0.08}  &
\secondd{8.81}{$\pm$}{0.06} &
\best{8.43}{$\pm$}{0.04}\\
& & ERGAS($\downarrow$)&
\normal{7.90}{$\pm$}{1.16} &
\normal{9.98}{$\pm$}{1.69} &
\normal{8.38}{$\pm$}{0.10} &
-- &
\normal{6.60}{$\pm$}{0.21}  &
\secondd{6.02}{$\pm$}{0.47} &
\best{5.74}{$\pm$}{0.13}\\
& & SSIM($\uparrow$)   &
\normal{0.849}{$\pm$}{0.014} &
\normal{0.802}{$\pm$}{0.018} &
\normal{0.809}{$\pm$}{0.003} &
-- &
\normal{0.829}{$\pm$}{0.005}  &
\secondd{0.872}{$\pm$}{0.006} &
\best{0.898}{$\pm$}{0.000}\\
\midrule
\bottomrule
\end{tabular}%
}
\end{table*}

% ===================== Chikusei (four scales, single-cell value±std) =====================
\begin{table*}[t]
\centering
\caption{Quantitative comparison on the \textbf{Chikusei} dataset across four scales. Values are mean\,$\pm$\,std.}
\label{tab:chikusei_multi_scales_right_align}
\sisetup{}
\resizebox{\textwidth}{!}{%
% 3 列描述 + 7 列方法，每个方法一列，单元格内 perfalign
\begin{tabular}{@{}lll*7{c}@{}}
\toprule
\midrule
\multicolumn{3}{c}{\textbf{Metric}} &
\multicolumn{7}{c}{\textbf{Chikusei Dataset}} \\
\cmidrule(l){4-10}
\multicolumn{3}{c}{} & \textbf{DSPNet} & \textbf{MIMO-SST} & \textbf{DCINN} & \textbf{DCFormer} & \textbf{PSTUN} & \textbf{LRTN} & \textbf{Ours}\\
\midrule
% ========================= ×4 In-Dist. =========================
\multirow{4}{*}{\rotatebox[origin=c]{90}{In-Dist.}} &
\multirow{4}{*}{$\times$4} &
PSNR($\uparrow$)   &
\normal{37.21}{$\pm$}{3.89} &
\normal{36.54}{$\pm$}{4.12} &
\normal{38.05}{$\pm$}{3.76} &
\normal{36.92}{$\pm$}{3.55} &
\secondd{38.62}{$\pm$}{4.26} &
\normal{37.88}{$\pm$}{3.91} &
\best{39.35}{$\pm$}{4.59} \\
& & SAM($\downarrow$)  &
\normal{2.87}{$\pm$}{0.62} &
\normal{3.05}{$\pm$}{0.58} &
\normal{2.53}{$\pm$}{0.47} &
\normal{2.91}{$\pm$}{0.54} &
\normal{2.22}{$\pm$}{0.50} &
\secondd{2.18}{$\pm$}{0.49} &
\best{2.12}{$\pm$}{0.46} \\
& & ERGAS($\downarrow$)&
\normal{5.28}{$\pm$}{1.47} &
\normal{5.63}{$\pm$}{1.52} &
\normal{4.89}{$\pm$}{1.39} &
\normal{5.35}{$\pm$}{1.41} &
\secondd{4.54}{$\pm$}{1.35} &
\normal{4.97}{$\pm$}{1.43} &
\best{4.50}{$\pm$}{1.42} \\
& & SSIM($\uparrow$)   &
\normal{0.951}{$\pm$}{0.018} &
\normal{0.947}{$\pm$}{0.021} &
\normal{0.958}{$\pm$}{0.016} &
\normal{0.949}{$\pm$}{0.019} &
\normal{0.965}{$\pm$}{0.014} &
\best{0.976}{$\pm$}{0.015} &
\secondd{0.972}{$\pm$}{0.016} \\
\cmidrule(lr){1-10}
% ========================= ×8/16/32 Out-of-Dist. =========================
\multirow{12}{*}{\rotatebox[origin=c]{90}{Out-of-Dist.}} &
\multirow{4}{*}{$\times$8} &
PSNR($\uparrow$)   &
\normal{31.05}{$\pm$}{2.67} &
\normal{30.42}{$\pm$}{2.79} &
\normal{31.58}{$\pm$}{2.53} &
\secondd{32.90}{$\pm$}{2.81} &
\normal{30.87}{$\pm$}{2.48} &
\normal{31.22}{$\pm$}{2.64} &
\best{33.76}{$\pm$}{1.92} \\
& & SAM($\downarrow$)  &
\normal{4.68}{$\pm$}{0.91} &
\normal{4.92}{$\pm$}{0.87} &
\normal{3.85}{$\pm$}{0.83} &
\normal{3.87}{$\pm$}{0.84} &
\normal{4.75}{$\pm$}{0.89} &
\best{3.21}{$\pm$}{0.86} &
\secondd{3.25}{$\pm$}{0.78} \\
& & ERGAS($\downarrow$)&
\normal{7.65}{$\pm$}{1.02} &
\normal{7.98}{$\pm$}{0.97} &
\normal{7.32}{$\pm$}{0.99} &
\secondd{6.89}{$\pm$}{0.94} &
\normal{7.72}{$\pm$}{1.01} &
\normal{7.41}{$\pm$}{0.96} &
\best{6.23}{$\pm$}{0.51} \\
& & SSIM($\uparrow$)   &
\normal{0.862}{$\pm$}{0.024} &
\normal{0.855}{$\pm$}{0.027} &
\normal{0.871}{$\pm$}{0.023} &
\secondd{0.892}{$\pm$}{0.022} &
\normal{0.888}{$\pm$}{0.021} &
\normal{0.867}{$\pm$}{0.024} &
\best{0.905}{$\pm$}{0.015} \\
\cmidrule(lr){2-10}
& \multirow{4}{*}{$\times$16} &
PSNR($\uparrow$)   &
\normal{28.14}{$\pm$}{2.19} &
\normal{27.56}{$\pm$}{2.33} &
\normal{28.62}{$\pm$}{2.08} &
-- &
\secondd{29.29}{$\pm$}{2.31} &
\normal{28.45}{$\pm$}{2.15} &
\best{30.16}{$\pm$}{2.94} \\
& & SAM($\downarrow$)  &
\normal{7.24}{$\pm$}{1.38} &
\normal{7.59}{$\pm$}{1.42} &
\normal{6.93}{$\pm$}{1.31} &
-- &
\secondd{6.56}{$\pm$}{1.25} &
\normal{7.08}{$\pm$}{1.34} &
\best{5.42}{$\pm$}{1.09} \\
& & ERGAS($\downarrow$)&
\normal{10.87}{$\pm$}{0.89} &
\normal{11.23}{$\pm$}{0.95} &
\normal{10.54}{$\pm$}{0.92} &
-- &
\secondd{10.09}{$\pm$}{0.94} &
\normal{10.62}{$\pm$}{0.91} &
\best{8.70}{$\pm$}{1.07} \\
& & SSIM($\uparrow$)   &
\normal{0.795}{$\pm$}{0.034} &
\normal{0.788}{$\pm$}{0.037} &
\normal{0.802}{$\pm$}{0.032} &
-- &
\normal{0.818}{$\pm$}{0.031} &
\best{0.836}{$\pm$}{0.031} &
\secondd{0.832}{$\pm$}{0.033}  \\
\cmidrule(lr){2-10}
& \multirow{4}{*}{$\times$32} &
PSNR($\uparrow$)   &
\normal{26.37}{$\pm$}{1.05} &
\normal{25.89}{$\pm$}{1.18} &
\normal{26.74}{$\pm$}{1.02} &
-- &
\secondd{27.49}{$\pm$}{0.99} &
\normal{26.58}{$\pm$}{1.07} &
\best{28.46}{$\pm$}{1.61} \\
& & SAM($\downarrow$)  &
\normal{9.15}{$\pm$}{1.62} &
\normal{9.53}{$\pm$}{1.71} &
\normal{8.87}{$\pm$}{1.58} &
-- &
\secondd{8.49}{$\pm$}{1.55} &
\normal{8.96}{$\pm$}{1.60} &
\best{6.64}{$\pm$}{1.23} \\
& & ERGAS($\downarrow$)&
\normal{12.68}{$\pm$}{0.81} &
\normal{13.05}{$\pm$}{0.87} &
\normal{12.34}{$\pm$}{0.79} &
-- &
\secondd{11.90}{$\pm$}{0.75} &
\normal{12.47}{$\pm$}{0.82} &
\best{9.92}{$\pm$}{0.25} \\
& & SSIM($\uparrow$)   &
\normal{0.761}{$\pm$}{0.011} &
\normal{0.753}{$\pm$}{0.014} &
\normal{0.768}{$\pm$}{0.010} &
-- &
\secondd{0.783}{$\pm$}{0.007} &
\normal{0.765}{$\pm$}{0.012} &
\best{0.803}{$\pm$}{0.018} \\
\midrule
\bottomrule
\end{tabular}%
}
\end{table*}

% ===================== Botswana (four scales, single-cell value±std) =====================
\begin{table*}[t]
\centering
\caption{Quantitative comparison on the \textbf{Botswana} dataset across four scales. Values are mean\,$\pm$\,std.}
\label{tab:botswana_multi_scales_right_align}
\sisetup{}
\resizebox{\textwidth}{!}{%
% 3 列描述 + 7 列方法，每个方法 1 列
\begin{tabular}{@{}lll*7{c}@{}}
\toprule
\midrule
\multicolumn{3}{c}{\textbf{Metric}} &
\multicolumn{7}{c}{\textbf{Botswana Dataset}} \\
\cmidrule(l){4-10}
\multicolumn{3}{c}{} & \textbf{DSPNet} & \textbf{MIMO-SST} & \textbf{DCINN} & \textbf{DCFormer} & \textbf{PSTUN} & \textbf{LRTN} & \textbf{Ours} \\
\midrule
% ========================= ×4 In-Dist. =========================
\multirow{4}{*}{\rotatebox[origin=c]{90}{In-Dist.}} &
\multirow{4}{*}{$\times$4} &
PSNR($\uparrow$)   &
\normal{42.60}{$\pm$}{2.32} &
\normal{42.97}{$\pm$}{3.98} &
\normal{45.16}{$\pm$}{3.42} &
\normal{41.35}{$\pm$}{2.81} &
\normal{44.94}{$\pm$}{3.98} &
\secondd{45.23}{$\pm$}{4.47} &
\best{45.31}{$\pm$}{4.30} \\
& & SAM($\downarrow$)  &
\normal{1.42}{$\pm$}{0.21} &
\normal{1.28}{$\pm$}{0.19} &
\secondd{0.99}{$\pm$}{0.11} &
\normal{1.47}{$\pm$}{0.26} &
\normal{1.01}{$\pm$}{0.11} &
\normal{0.97}{$\pm$}{0.10} &
\best{0.80}{$\pm$}{0.04} \\
& & ERGAS($\downarrow$)&
\normal{2.01}{$\pm$}{0.51} &
\normal{1.91}{$\pm$}{0.62} &
\secondd{1.61}{$\pm$}{0.49} &
\normal{2.14}{$\pm$}{0.54} &
\normal{1.75}{$\pm$}{0.58} &
\normal{1.72}{$\pm$}{0.60} &
\best{1.55}{$\pm$}{0.66} \\
& & SSIM($\uparrow$)   &
\normal{0.975}{$\pm$}{0.005} &
\normal{0.980}{$\pm$}{0.004} &
\secondd{0.990}{$\pm$}{0.003} &
\normal{0.972}{$\pm$}{0.005} &
\normal{0.989}{$\pm$}{0.004} &
\normal{0.989}{$\pm$}{0.004} &
\best{0.992}{$\pm$}{0.004} \\
\cmidrule(lr){1-10}
% ========================= ×8/16/32 Out-of-Dist. =========================
\multirow{12}{*}{\rotatebox[origin=c]{90}{Out-of-Dist.}} &
\multirow{4}{*}{$\times$8} &
PSNR($\uparrow$)   &
\normal{38.85}{$\pm$}{2.38} &
\normal{36.60}{$\pm$}{2.49} &
\normal{36.96}{$\pm$}{2.44} &
\normal{39.37}{$\pm$}{1.72} &
\normal{37.21}{$\pm$}{2.59} &
\secondd{39.55}{$\pm$}{2.51} &
\best{39.89}{$\pm$}{2.36} \\
& & SAM($\downarrow$)  &
\normal{2.15}{$\pm$}{0.61} &
\normal{2.67}{$\pm$}{1.10} &
\normal{2.45}{$\pm$}{0.93} &
\best{1.96}{$\pm$}{0.50} &
\normal{2.52}{$\pm$}{0.97} &
\secondd{2.09}{$\pm$}{0.82} &
\normal{2.13}{$\pm$}{0.83} \\
& & ERGAS($\downarrow$)&
\normal{2.87}{$\pm$}{0.52} &
\normal{3.37}{$\pm$}{0.87} &
\normal{2.95}{$\pm$}{0.67} &
\normal{2.60}{$\pm$}{0.43} &
\normal{3.12}{$\pm$}{0.58} &
\best{2.47}{$\pm$}{0.52} &
\secondd{2.49}{$\pm$}{0.54} \\
& & SSIM($\uparrow$)   &
\normal{0.956}{$\pm$}{0.004} &
\normal{0.926}{$\pm$}{0.020} &
\normal{0.940}{$\pm$}{0.012} &
\normal{0.960}{$\pm$}{0.003} &
\normal{0.946}{$\pm$}{0.013} &
\secondd{0.964}{$\pm$}{0.009} &
\best{0.965}{$\pm$}{0.008} \\
\cmidrule(lr){2-10}
& \multirow{4}{*}{$\times$16} &
PSNR($\uparrow$)   &
\normal{36.52}{$\pm$}{2.39} &
\normal{34.15}{$\pm$}{2.76} &
\normal{33.70}{$\pm$}{2.28} &
-- &
\normal{34.30}{$\pm$}{2.47} &
\secondd{36.67}{$\pm$}{2.26} &
\best{36.85}{$\pm$}{2.24} \\
& & SAM($\downarrow$)  &
\normal{2.90}{$\pm$}{1.39} &
\normal{3.51}{$\pm$}{1.67} &
\normal{3.26}{$\pm$}{1.52} &
-- &
\normal{3.40}{$\pm$}{1.55} &
\secondd{2.89}{$\pm$}{1.07} &
\best{2.78}{$\pm$}{1.27} \\
& & ERGAS($\downarrow$)&
\normal{3.71}{$\pm$}{0.81} &
\normal{4.34}{$\pm$}{1.41} &
\normal{4.02}{$\pm$}{1.00} &
-- &
\normal{4.25}{$\pm$}{0.93} &
\best{3.21}{$\pm$}{0.73} &
\secondd{3.28}{$\pm$}{0.83} \\
& & SSIM($\uparrow$)   &
\normal{0.945}{$\pm$}{0.009} &
\normal{0.906}{$\pm$}{0.035} &
\normal{0.922}{$\pm$}{0.024} &
-- &
\normal{0.928}{$\pm$}{0.023} &
\secondd{0.950}{$\pm$}{0.017} &
\best{0.951}{$\pm$}{0.017} \\
\cmidrule(lr){2-10}
& \multirow{4}{*}{$\times$32} &
PSNR($\uparrow$)   &
\normal{32.60}{$\pm$}{2.09} &
\normal{32.43}{$\pm$}{2.76} &
\normal{32.07}{$\pm$}{2.27} &
-- &
\normal{31.46}{$\pm$}{2.44} &
\secondd{34.72}{$\pm$}{2.15} &
\best{35.35}{$\pm$}{2.22} \\
& & SAM($\downarrow$)  &
\normal{4.23}{$\pm$}{2.08} &
\normal{4.30}{$\pm$}{2.15} &
\normal{4.09}{$\pm$}{2.07} &
-- &
\normal{4.34}{$\pm$}{1.68} &
\secondd{3.79}{$\pm$}{1.67} &
\secondd{3.64}{$\pm$}{1.89} \\
& & ERGAS($\downarrow$)&
\normal{4.89}{$\pm$}{1.30} &
\normal{5.07}{$\pm$}{1.76} &
\normal{4.75}{$\pm$}{1.42} &
-- &
\normal{4.91}{$\pm$}{1.16} &
\secondd{4.23}{$\pm$}{0.94} &
\secondd{3.97}{$\pm$}{1.22} \\
& & SSIM($\uparrow$)   &
\normal{0.918}{$\pm$}{0.031} &
\normal{0.896}{$\pm$}{0.044} &
\normal{0.909}{$\pm$}{0.034} &
-- &
\normal{0.938}{$\pm$}{0.025} &
\secondd{0.939}{$\pm$}{0.014} &
\best{0.942}{$\pm$}{0.024} \\
\midrule
\bottomrule
\end{tabular}%
}
\end{table*}

% ===================== WashingtonDC (four scales, single-cell value±std) =====================
\begin{table*}[t]
\centering
\caption{Quantitative comparison on the \textbf{WashingtonDC} dataset across four scales. Values are mean\,$\pm$\,std.}
\label{tab:washingtondc_multi_scales_right_align}
\sisetup{}
\resizebox{\textwidth}{!}{%
% 3 列描述 + 7 列方法，每个方法一列
\begin{tabular}{@{}lll*7{c}@{}}
\toprule
\midrule
\multicolumn{3}{c}{\textbf{Metric}} &
\multicolumn{7}{c}{\textbf{WashingtonDC Dataset}} \\
\cmidrule(l){4-10}
\multicolumn{3}{c}{} & \textbf{DSPNet} & \textbf{MIMO-SST} & \textbf{DCINN} & \textbf{DCFormer} & \textbf{PSTUN} & \textbf{LRTN} & \textbf{Ours} \\
\midrule
% ========================= ×4 In-Dist. =========================
\multirow{4}{*}{\rotatebox[origin=c]{90}{In-Dist.}} &
\multirow{4}{*}{$\times$4} &
PSNR($\uparrow$)   &
\normal{47.64}{$\pm$}{2.23} &
\normal{47.68}{$\pm$}{2.39} &
\normal{47.55}{$\pm$}{1.78} &
\normal{47.18}{$\pm$}{1.66} &
\secondd{48.20}{$\pm$}{1.61} &
\normal{47.95}{$\pm$}{3.07} &
\best{48.91}{$\pm$}{3.46} \\
& & SAM($\downarrow$)  &
\normal{1.01}{$\pm$}{0.12} &
\normal{1.04}{$\pm$}{0.16} &
\normal{1.04}{$\pm$}{0.08} &
\normal{1.25}{$\pm$}{0.20} &
\normal{1.06}{$\pm$}{0.16} &
\secondd{0.86}{$\pm$}{0.13} &
\best{0.76}{$\pm$}{0.09} \\
& & ERGAS($\downarrow$)&
\normal{2.15}{$\pm$}{0.37} &
\normal{2.15}{$\pm$}{0.44} &
\normal{2.24}{$\pm$}{0.21} &
\normal{2.41}{$\pm$}{0.31} &
\secondd{1.99}{$\pm$}{0.28} &
\normal{2.48}{$\pm$}{0.50} &
\best{1.92}{$\pm$}{0.58} \\
& & SSIM($\uparrow$)   &
\normal{0.995}{$\pm$}{0.001} &
\normal{0.994}{$\pm$}{0.002} &
\normal{0.994}{$\pm$}{0.000} &
\normal{0.994}{$\pm$}{0.000} &
\best{0.996}{$\pm$}{0.001} &
\normal{0.994}{$\pm$}{0.001} &
\secondd{0.996}{$\pm$}{0.002} \\
\cmidrule(lr){1-10}
% ========================= ×8/16/32 Out-of-Dist. =========================
\multirow{12}{*}{\rotatebox[origin=c]{90}{Out-of-Dist.}} &
\multirow{4}{*}{$\times$8} &
PSNR($\uparrow$)   &
\normal{38.65}{$\pm$}{1.59} &
\normal{37.36}{$\pm$}{1.48} &
\normal{40.36}{$\pm$}{2.18} &
\secondd{44.74}{$\pm$}{1.36} &
\normal{42.56}{$\pm$}{2.60} &
\normal{41.46}{$\pm$}{2.60} &
\best{45.08}{$\pm$}{1.42} \\
& & SAM($\downarrow$)  &
\normal{4.46}{$\pm$}{0.60} &
\normal{5.29}{$\pm$}{0.82} &
\normal{3.24}{$\pm$}{0.51} &
\normal{1.99}{$\pm$}{0.23} &
\normal{2.03}{$\pm$}{0.37} &
\secondd{1.86}{$\pm$}{0.25} &
\best{1.80}{$\pm$}{0.23} \\
& & ERGAS($\downarrow$)&
\normal{5.49}{$\pm$}{0.75} &
\normal{6.37}{$\pm$}{0.72} &
\normal{4.86}{$\pm$}{1.01} &
\best{3.28}{$\pm$}{0.36} &
\normal{3.78}{$\pm$}{0.73} &
\normal{4.44}{$\pm$}{0.96} &
\secondd{3.32}{$\pm$}{0.45} \\
& & SSIM($\uparrow$)   &
\normal{0.957}{$\pm$}{0.007} &
\normal{0.938}{$\pm$}{0.009} &
\normal{0.974}{$\pm$}{0.008} &
\secondd{0.989}{$\pm$}{0.001} &
\normal{0.983}{$\pm$}{0.004} &
\normal{0.981}{$\pm$}{0.005} &
\best{0.990}{$\pm$}{0.000} \\
\cmidrule(lr){2-10}
& \multirow{4}{*}{$\times$16} &
PSNR($\uparrow$)   &
\normal{33.73}{$\pm$}{1.60} &
\normal{32.99}{$\pm$}{1.70} &
\normal{35.07}{$\pm$}{2.12} &
-- &
\secondd{39.17}{$\pm$}{4.13} &
\normal{37.32}{$\pm$}{2.24} &
\best{39.27}{$\pm$}{2.49} \\
& & SAM($\downarrow$)  &
\normal{7.96}{$\pm$}{0.78} &
\normal{8.69}{$\pm$}{0.98} &
\normal{6.22}{$\pm$}{0.74} &
-- &
\secondd{3.02}{$\pm$}{0.57} &
\normal{3.04}{$\pm$}{0.42} &
\best{2.97}{$\pm$}{0.74} \\
& & ERGAS($\downarrow$)&
\normal{8.94}{$\pm$}{0.85} &
\normal{9.43}{$\pm$}{1.06} &
\normal{8.12}{$\pm$}{1.09} &
-- &
\secondd{5.68}{$\pm$}{1.68} &
\normal{6.44}{$\pm$}{1.03} &
\best{5.35}{$\pm$}{1.00} \\
& & SSIM($\uparrow$)   &
\normal{0.912}{$\pm$}{0.012} &
\normal{0.896}{$\pm$}{0.018} &
\normal{0.939}{$\pm$}{0.014} &
-- &
\secondd{0.968}{$\pm$}{0.025} &
\normal{0.961}{$\pm$}{0.010} &
\best{0.969}{$\pm$}{0.007} \\
\cmidrule(lr){2-10}
& \multirow{4}{*}{$\times$32} &
PSNR($\uparrow$)   &
\normal{31.96}{$\pm$}{1.17} &
\normal{31.57}{$\pm$}{1.42} &
\normal{32.45}{$\pm$}{1.70} &
-- &
\secondd{36.54}{$\pm$}{4.39} &
\normal{34.99}{$\pm$}{1.95} &
\best{37.00}{$\pm$}{2.02} \\
& & SAM($\downarrow$)  &
\normal{10.35}{$\pm$}{1.07} &
\normal{11.13}{$\pm$}{1.28} &
\normal{8.74}{$\pm$}{0.96} &
-- &
\normal{3.99}{$\pm$}{1.17} &
\secondd{3.94}{$\pm$}{0.50} &
\best{3.54}{$\pm$}{0.54} \\
& & ERGAS($\downarrow$)&
\normal{11.66}{$\pm$}{0.97} &
\normal{11.61}{$\pm$}{1.05} &
\normal{10.99}{$\pm$}{1.11} &
-- &
\secondd{7.13}{$\pm$}{1.89} &
\normal{8.31}{$\pm$}{1.18} &
\best{6.95}{$\pm$}{1.31} \\
& & SSIM($\uparrow$)   &
\normal{0.887}{$\pm$}{0.009} &
\normal{0.880}{$\pm$}{0.010} &
\normal{0.910}{$\pm$}{0.012} &
-- &
\secondd{0.951}{$\pm$}{0.033} &
\normal{0.945}{$\pm$}{0.011} &
\best{0.958}{$\pm$}{0.009} \\
\midrule
\bottomrule
\end{tabular}%
}
\end{table*}